\documentclass[twocolumn]{emulateapj}

\usepackage{graphicx}
\usepackage{epsfig}

\newcommand\be{\begin{equation}}
\newcommand\ba{\begin{eqnarray}}
\newcommand\ee{\end{equation}}
\newcommand\ea{\end{eqnarray}}

\newcommand\nn{\nonumber}

\shorttitle{PHOTON RINGS AROUND KERR AND KERR-LIKE BLACK HOLES}
\shortauthors{JOHANNSEN}

\begin{document}

\title{PHOTON RINGS AROUND KERR AND KERR-LIKE BLACK HOLES}

\author{Tim Johannsen\footnote{CITA National Fellow}}
\affil{Department of Physics and Astronomy, University of Waterloo, 200 University Avenue West, Waterloo, ON N2L 3G1, Canada \\
Canadian Institute for Theoretical Astrophysics, University of Toronto, Toronto, ON M5S 3H8, Canada \\
Perimeter Institute for Theoretical Physics, 31 Caroline Street North, Waterloo, ON N2L 2Y5, Canada}

\begin{abstract}

Very-long baseline interferometric observations have resolved structure on scales of only a few Schwarzschild radii around the supermassive black holes at the centers of our Galaxy and M87. In the near future, such observations are expected to image the shadows of these black holes together with a bright and narrow ring surrounding their shadows. For a Kerr black hole, the shape of this photon ring is nearly circular unless the black hole spins very rapidly. Whether or not, however, astrophysical black holes are truly described by the Kerr metric as encapsulated in the no-hair theorem still remains an untested assumption. For black holes that differ from Kerr black holes, photon rings have been shown numerically to be asymmetric for small to intermediate spins. In this paper, I calculate semi-analytic expressions of the shapes of photon rings around black holes described by a new Kerr-like metric which is valid for all spins. I show that photon rings in this spacetime are affected by two types of deviations from the Kerr metric which can cause the ring shape to be highly asymmetric. I argue that the ring asymmetry is a direct measure of a potential violation of the no-hair theorem and that both types of deviations can be detected independently if the mass and distance of the black hole are known. In addition, I obtain approximate expressions of the diameters, displacements, and asymmetries of photon rings around Kerr and Kerr-like black holes.

\end{abstract}


\section{INTRODUCTION}

The supermassive black holes at the centers of the Milky Way and M87 are the prime targets for high-resolution very-long baseline interferometric (VLBI) observations with the Event Horizon Telescope (EHT), a planned global array of (sub)millimeter telescopes (Doeleman et al. 2009a,b; Fish et al. 2009). Recent VLBI observations at 230~GHz with a three-station array consisting of the James Clerk Maxwell Telescope (JCMT) and Sub-Millimeter Array (SMA) in Hawaii, the Submillimeter Telescope Observatory (SMTO) in Arizona, and several dishes of the Combined Array for Research in Millimeter-wave Astronomy (CARMA) in California resolved structure on scales of only $4r_S$ (Sgr~A*; Doeleman et al. 2008) and $5.5r_S$ (M87; Doeleman et al. 2012), where $r_S\equiv2GM/c^2$ is the Schwarzschild radius of a black hole with mass $M$ and where $G$ and $c$ are the gravitational constant and the speed of light, respectively. Similar observations also detected time variability on these scales in Sgr~A* (Fish et al. 2011). Such measurements have demonstrated the feasibility of VLBI imaging of Sgr~A* and the supermassive black hole in M87 on (sub-)event horizon scales. Simulations based on larger telescope arrays support the possibility of directly imaging the shadows of these two supermassive black holes with the EHT (Doeleman et al. 2009a).

Using the measurements of the mass and distance of Sgr~A* from the monitoring of close stellar orbits (Ghez et al. 2008; Gillessen et al. 2009), Broderick et al. (2009a, 2011a) combined the VLBI observations of Sgr~A* with measurements of its spectral energy distribution (Yuan et al. 2004 and references therein; Marrone 2006) and obtained values of the spin magnitude and direction employing a radiatively inefficient accretion flow (RIAF) model (Yuan et al. 2003; Broderick \& Loeb 2006a). Likewise, Huang et al. (2009) analyzed the VLBI data of Sgr~A* using an accretion flow model with plasma wave heating for several different values of the spin and disk inclination. Mo\'scibrodzka et al. (2009) and Dexter et al. (2009, 2010) fitted the VLBI and spectral data to sets of images obtained from three-dimensional general-relativistic magnetohydrodynamic (3D-GRMHD) simulations (Gammie et al. 2003; Fragile et al. 2007, 2009; McKinney \& Blandford 2009). In the future, the determination of the spin and orientation of the black hole can be further improved with a multiwavelength study of polarization (Broderick \& Loeb 2006b; see, also, Schnittman \& Krolik 2009, 2010).

The shadow is a prominent feature of resolved accretion flow images of supermassive black holes. The shape of the shadow depends uniquely on the mass, spin, and inclination of the black hole (e.g., Falcke et al. 2000) as well as on potential deviations from the Kerr metric (Johannsen \& Psaltis 2010b; hereafter JP10b). Images of optically thin accretion flows also reveal a characteristic bright and narrow ring which surrounds the shadow, the so-called photon ring. This ring is the projection along null geodesics of photon trajectories that wind around the black hole many times. Thanks to the long path length through the emitting medium of the photons that comprise the ring, these photons can make a significantly larger contribution to the observed flux than individual photons outside of the ring. Images of the shadows of black holes with optically and geometrically thin accretion disks and of their photon rings have been calculated by several authors (Bardeen 1973; Cunningham \& Bardeen 1973; Luminet 1979; Sikora 1979; Fukue \& Yokoyama 1988; Jaroszy\'nski et al. 1992; Viergutz 1993; Falcke et al. 2000; Takahashi 2004; Beckwith \& Done 2005; Bozza 2008; JP10b; Chan et al. 2013). This photon ring is clearly visible in several 3D-GRMHD simulations reported to date (Mo\'scibrodzka et al. 2009; Dexter et al. 2009; Shcherbakov \& Penna 2010).

The photon rings of Sgr~A* and the supermassive black hole in M87 have angular sizes of $\sim53~{\rm \mu as}$ and $\sim22~{\rm \mu as}$, respectively, corresponding to a diameter of $\sim5.2r_S$ (Johannsen et al. 2012). Since the shape of the photon ring of a black hole is determined only by the geometry of the underlying spacetime, it is independent of the complicated structure of the accretion flow making it an excellent target of VLBI imaging observations. For a Schwarzschild black hole, the photon ring is circular and centered on the black hole. For Kerr black holes with a spin $0\leq |a|\leq GM/c^2$, where $a\equiv GJ/cM$ is the spin parameter of a black hole with angular momentum $J$, the photon ring retains a nearly circular shape and is displaced off center (Takahashi 2004; Beckwith \& Done 2005; JP10b). The shape of the photon ring around Kerr black holes becomes asymmetric only in the case of extremely high spin values $a\gtrsim0.9GM/c^2$ if the inclination angle is large (JP10b; Chan et al. 2013). Since the ring diameter of a Kerr black hole depends primarily on the black hole mass, imaging the ring around Sgr~A* with VLBI can improve its current mass measurement in conjunction with the monitoring of the ephemerides of stars in its vicinity (Johannsen et al. 2012).

Images of photon rings can be significantly altered if the black hole is not described by the Kerr metric (JP10b). According to the no-hair theorem, the Kerr metric is the unique black hole metric in general relativity and black holes are fully characterized by their masses and spins (see, e.g., Heusler 1996). However, while there exists evidence for the presence of event horizons in astrophysical black holes (e.g., Narayan et al. 1997; Broderick et al. 2009b), the Kerr nature of black holes still remains untested to date. Several approaches have been developed to test the no-hair theorem which are based on Kerr-like metrics that deviate from the Kerr spacetime in parametric form (e.g., Manko \& Novikov 1992; Collins \& Hughes 2004; Glampedakis \& Babak 2006; Vigeland \& Hughes 2010; Johannsen \& Psaltis 2011b; Vigeland et al. 2011). Observational signatures of potential violations of the no-hair theorem can then be studied in either the electromagnetic (e.g., Johannsen \& Psaltis 2010a,b, 2011a, 2013; Bambi \& Barausse 2011; Bambi 2012a,b, 2013a,b; Krawczynski 2012) or gravitational wave spectra (see Gair et al. 2013; Yunes \& Siemens 2013 for reviews).

JP10b analyzed the shapes of photon rings of compact objects described by a quasi-Kerr metric (Glampedakis \& Babak 2006). This metric contains a quadrupole moment of the form
\be
Q=-M\left[ a^2+\epsilon \left(\frac{GM}{c^2}\right)^2 \right],
\label{eq:QQK}
\ee
where the parameter $\epsilon$ is dimensionless and measures potential deviations from the Kerr metric. When $\epsilon=0$, the quasi-Kerr metric reduces to the familiar Kerr metric. For nonzero values of the parameter $\epsilon$, photon rings acquire an asymmetric shape. Therefore, the ring shape can be used as a direct measure of a potential violation of the no-hair theorem (JP10b).

By construction, however, the quasi-Kerr metric is only appropriate for the description of black holes that do not spin rapidly (Glampedakis \& Babak 2006; Johannsen \& Psaltis 2010a; Johannsen 2013a). In order to study deviations from the Kerr metric for black holes with arbitrary spins $|a|\leq GM/c^2$, Johannsen \& Psaltis (2011b) constructed a Kerr-like metric that can be used to model accretion flows even around rapidly-spinning black holes. Similar to the quasi-Kerr metric, this metric depends on one set of free parameters in addition to the mass and spin of the compact object. While this structure is adequate for the study of deviations from the Kerr metric of a certain type, it is not the most general form to parameterize deviations from the Kerr metric, because such a Kerr-like metric should depend on at least four independent functions that measure these deviations instead of only one. Furthermore, as in the quasi-Kerr metric, the shapes of the shadows and photon rings of black holes in this metric can only be calculated numerically, because geodesic orbits are generally not integrable. Images of photon rings and accretion flows around black holes and some exotic objects in other theories of gravity were also analyzed by Bambi \& Yoshida (2010), Amarilla et al. (2010), Bambi et al. (2012), Amarilla \& Eiroa (2012), and Bambi (2013b). Chen \& Jing (2012) and Liu et al. (2012a) studied the strong gravitational lensing near Kerr-like compact objects.

In this paper, I study the shapes of photon rings around black holes that are described by a new Kerr-like metric proposed recently (Johannsen 2013b). This metric depends on four sets of deviation functions and describes a black hole for all values of the spin $|a|\leq GM/c^2$. As in the case of the Kerr metric, the equations of motion of particles on geodesic orbits in this metric are fully integrable thanks to the existence of three independent constants of motion. This property allows for the calculation in closed form of the shapes of photon rings in the image plane of an observer at a large distance from the black hole in the limit of vanishing ring thickness. In this case, the photon ring reduces to the projection along null geodesics of the circular photon orbit in the equatorial plane of the black hole, which delineates the silhouette of the shadow of the black hole if its accretion flow is optically thin.

I derive semi-analytic expressions of such photon rings around the Kerr-like black holes described by this metric and of the positions and shapes of these rings in the image plane. I show that the location of the circular photon orbit in the equatorial plane of the black hole as well as the shape of the photon ring depend on the mass, spin, and inclination of the black hole and on two types of deviations from the Kerr metric. In addition, I obtain approximate expressions of the ring diameter, displacement, and asymmetry in terms of these parameters for Kerr and Kerr-like black holes from least-squares fits of a large set of calculated photon ring images.

In Section~\ref{sec:metric}, I summarize the Kerr and Kerr-like metrics that I use in this paper as well as some of their properties. In Section~\ref{sec:rings}, I determine the location of the circular photon orbit and calculate expressions of the shapes of photon rings in the image plane. In Section~\ref{sec:DDA}, I derive expressions of the diameters, displacements, and asymmetries of these photon rings and find approximate fit formulae. I discuss my conclusions in Section~\ref{sec:conclusions}. From here on, I use geometric units and set $G=c=1$. In these units, length (and time) scales are expressed in terms of the mass of the black hole, where $M=GM/c^2=r_S/2$.

\section{KERR AND KERR-LIKE BLACK HOLE METRICS}
\label{sec:metric}

In Boyer-Lindquist coordinates, the Kerr metric $g_{\mu\nu}^{\rm K}$ is given by the elements
\ba
g_{tt}^{\rm K} &=&-\left(1-\frac{2Mr}{\Sigma}\right), \nonumber \\
g_{t\phi}^{\rm K} &=& -\frac{2Mar\sin^2\theta}{\Sigma}, \nonumber \\
g_{rr}^{\rm K} &=& \frac{\Sigma}{\Delta}, \nonumber \\
g_{\theta \theta}^{\rm K} &=& \Sigma, \nonumber \\
g_{\phi \phi}^{\rm K} &=& \left(r^2+a^2+\frac{2Ma^2r\sin^2\theta}{\Sigma}\right)\sin^2\theta,
\label{kerr}
\ea
where
\ba
\Delta &\equiv& r^2-2Mr+a^2, \nn \\
\Sigma &\equiv& r^2+a^2\cos^2 \theta.
\label{deltasigma}
\ea

The Kerr-like black hole metric constructed by Johannsen (2013b) is given by the elements
\ba
g_{tt} &=& -\frac{\tilde{\Sigma}[\Delta-a^2A_2(r)^2\sin^2\theta]}{[(r^2+a^2)A_1(r)-a^2A_2(r)\sin^2\theta]^2}, \nn \\
g_{t\phi} &=& -\frac{a[(r^2+a^2)A_1(r)A_2(r)-\Delta]\tilde{\Sigma}\sin^2\theta}{[(r^2+a^2)A_1(r)-a^2A_2(r)\sin^2\theta]^2}, \nn \\
g_{rr} &=& \frac{\tilde{\Sigma}}{\Delta A_5(r)}, \nn \\
g_{\theta \theta} &=& \tilde{\Sigma}, \nn \\
g_{\phi \phi} &=& \frac{\tilde{\Sigma} \sin^2 \theta \left[(r^2 + a^2)^2 A_1(r)^2 - a^2 \Delta \sin^2 \theta \right]}{[(r^2+a^2)A_1(r)-a^2A_2(r)\sin^2\theta]^2},
\label{eq:metric}
\ea
where
\ba
A_1(r) &=& 1 + \sum_{n=3}^\infty \alpha_{1n} \left( \frac{M}{r} \right)^n, 
\label{eq:A1}\\
A_2(r) &=& 1 + \sum_{n=2}^\infty \alpha_{2n} \left( \frac{M}{r} \right)^n, 
\label{eq:A2}\\
A_5(r) &=& 1 + \sum_{n=2}^\infty \alpha_{5n} \left( \frac{M}{r} \right)^n, 
\label{eq:A5}\\
\tilde{\Sigma} &=& \Sigma + f(r), 
\label{eq:Sigmatilde}\\
f(r) &=& \sum_{n=3}^\infty\epsilon_n \frac{M^n}{r^{n-2}}.
\label{eq:f}
\ea

The metric in Equation~(\ref{eq:metric}) contains four free functions, $f(r)$, $A_1(r)$, $A_2(r)$, and $A_5(r)$, each of which depends on a set of parameters which measure potential deviations from the Kerr metric. The Kerr metric is recovered when all deviation parameters vanish, i.e., when $f(r)=0$, $A_1(r)=A_2(r)=A_5(r)=1$. Thanks to these deviation functions, this metric can describe black holes in alternative theories of gravity that differ from Kerr black holes. However, instead of focusing on any particular modified gravity theory, which would a priori limit the study of non-Kerr black holes to include only those that are predicted by that theory, it was designed in a phenomenological approach, which encompasses large classes of modified theories of gravity corresponding to different choices of the deviation functions. Examples of such theories, which admit non-Kerr black holes that can be described by the metric in Equation~(\ref{eq:metric}) in the appropriate limits, include Chern-Simons gravity and Einstein-Dilaton-Gauss-Bonnet gravity (see Johannsen 2013b). Astrophysical observations can then be used to obtain constraints on deviations from the Kerr metric or to infer properties of the underlying, but so far unknown modified theory of gravity should the no-hair theorem be violated.

Just as the Kerr metric, this metric is stationary, axisymmetric, and asymptotically flat. Since all stationary, axisymmetric, asymptotically flat, vacuum metrics in general relativity can be written in terms of four independent functions (see, e.g., Wald 1984 for a detailed discussion), it is natural to introduce four deviation functions to the Kerr metric instead of only one as in the quasi-Kerr metric (Glampedakis \& Babak 2006) and the metric proposed by Johannsen \& Psaltis (2011b). While the latter two metrics are thus less general than the metric in Equation~(\ref{eq:metric}), they remain valid frameworks for the study of observational signatures of potential deviations from the Kerr metric.

The deviation functions in Equations~(\ref{eq:A1})--(\ref{eq:f}), in turn, are written as power series in $M/r$. Such a choice allows one to parameterize the deviations in a physically reasonable way given the enormous freedom of potential departures from the Kerr metric. The coefficients of these series are chosen so that the deviations they introduce are consistent with all current weak-field tests of general relativity such as those performed in the solar system and with double neutron stars (see Will 2006). Since one would expect that the lowest-order coefficients in these power series have the strongest effect on the observed signals, one could image to constrain these coefficients successively order by order with increasing accuracy of the measurements.

For these reasons, I will only study black holes and their photon rings in this paper in the lowest-order metric, i.e, a given black hole only depends on its mass $M$, spin $a$, and the deviation parameters $\epsilon_3$, $\alpha_{13}$, $\alpha_{22}$, and $\alpha_{52}$. In the next section I will show that in this metric photon rings only depend on two deviation parameters, $\alpha_{13}$ and $\alpha_{22}$, which modify the $(t,t)$, $(t,\phi)$, and $(\phi,\phi)$ components of the metric. Ultimately, constraints on the coefficients of higher order terms in the deviation functions will be required to obtain a deeper understanding of the full character of the deviation functions.

While the deviation parameters $\epsilon_3$, $\alpha_{13}$, $\alpha_{22}$, and $\alpha_{52}$ can be either positive or negative, they have to be larger than certain lower bounds, because otherwise the central object is a naked singularity and not a black hole. These bounds are given by the expressions (Johannsen 2013b)
\ba
\epsilon_3 &>& B_3,~~~~~\alpha_{13} > B_3, \nn \\
\alpha_{22} &>& B_2,~~~~~\alpha_{52} > B_2,
\ea
where
\ba
B_2 &\equiv& - \frac{\left(M+\sqrt{M^2-a^2}\right)^2}{M^2}, \nn \\
B_3 &\equiv& - \frac{\left(M+\sqrt{M^2-a^2}\right)^3}{M^3}.
\label{eq:lowerbounds}
\ea
Smaller values of these parameters are excluded. For all allowed values of the deviation parameters, the event horizon $r_+$ is independent of these parameters and coincides with the event horizon of a Kerr black hole,
\be
r_+ \equiv M+\sqrt{M^2-a^2}.
\ee

Thanks to the stationarity and axisymmetry of the metric in Equation~(\ref{eq:metric}), the energy $E$ and angular momentum $L_z$ about the $z$-axis of particles moving along geodesics are conserved. The Kerr metric possesses a third conserved constant of motion, the so-called Carter constant (Carter 1968). Similarly, the metric in Equation~(\ref{eq:metric}) admits a Carter constant of a similar form which is given by the expression (Johannsen 2013b)
\be
C \equiv p_\theta^2 - (L_z-aE)^2 + \mu^2a^2\cos^2\theta + \frac{1}{\sin^2\theta}(L_z-aE\sin^2\theta)^2,
\label{eq:CarterQ}
\ee
where $p_\theta/\tilde{\Sigma}=p^\theta$ is the $\theta$-component of the particle's four-momentum and $\mu$ is the rest mass of the particle. 

The existence of three constants of motion makes geodesic motion in both of these metrics fully integrable. This property greatly simplifies ray tracing algorithms in these spacetimes which are essential for the calculation of images of black hole accretion flows, because the equations of motion of particles and, in particular, of photons can be written in first order form (see, e.g., Rauch \& Blandford 1994; Dexter \& Agol 2009 for such algorithms in the Kerr metric).

\section{CIRCULAR PHOTON ORBIT AND PHOTON RING}
\label{sec:rings}

In this section, I study the location of the circular photon orbit in the equatorial plane of a black hole described by the metric in Equation~(\ref{eq:metric}) and its projection along null geodesics onto the image plane of a distant observer. The image of the circular photon orbit marks the silhouette of the black hole shadow (Bardeen 1973; Cunningham \& Bardeen 1973) which corresponds to the inner edge of the photon ring. Photon rings have a small thickness which is caused by the projection along null geodesics of photon paths that trace many orbits around the black hole but escape eventually. These curves are generally complicated functions of the radius $r$ and the polar angle $\theta$. In the equatorial plane, however, the trajectory is circular and located at a radius $r_{\rm ph}$, the radius of the circular photon orbit. In this paper, I will neglect the thickness of the photon ring and only discuss thin photon rings which are the projection along null geodesics of the circular photon orbit.

For a Schwarzschild black hole, the circular photon orbit is located at the radius $r_{\rm ph}=3M$. In the Kerr metric, this radius is a function of the spin of the black hole. As the spin increases, the circular photon orbit approaches the event horizon and merges with the coordinate radius of the event horizon when $a=M$. In the Kerr-like metric given by Equation~(\ref{eq:metric}), the location of the circular photon orbit also depends on the deviation parameters (c.f., Johannsen \& Psaltis 2010a, 2011b).

In order to calculate the location of the circular photon orbit in the metric given by Equation~(\ref{eq:metric}), I make use of the expressions of the energy $E$ and axial angular momentum $L_z$ of massive particles on circular equatorial orbits which I already calculated in Johannsen (2013b). In the limit of vanishing rest mass, the energy and axial angular momentum diverge at the location of the circular photon orbit. Both the energy and axial angular momentum for massive particles on circular equatorial orbits (including the ISCO, the innermost stable circular orbit) generally depend on the spin $a$ and the deviation parameters $\epsilon_3$, $\alpha_{13}$, and $\alpha_{22}$ but are independent of the deviation parameter $\alpha_{52}$. The latter deviation parameter only affects the $(r,r)$ component of the metric, which plays no role for particles on circular equatorial orbits in this regard (Johannsen 2013b). The location of the circular photon orbit, however, depends only on the spin and the deviation parameters $\alpha_{13}$ and $\alpha_{22}$, because all terms that are affected by the parameter $\epsilon_3$ also contain the rest mass of the particle which vanishes for photons. I, then, calculate the radius of the circular photon orbit by numerically finding the root of the inverse of the energy $E$. Equivalently, the axial angular momentum $L_z$ could be used instead, which would yield the same result.

\begin{figure}[ht]
\begin{center}
\psfig{figure=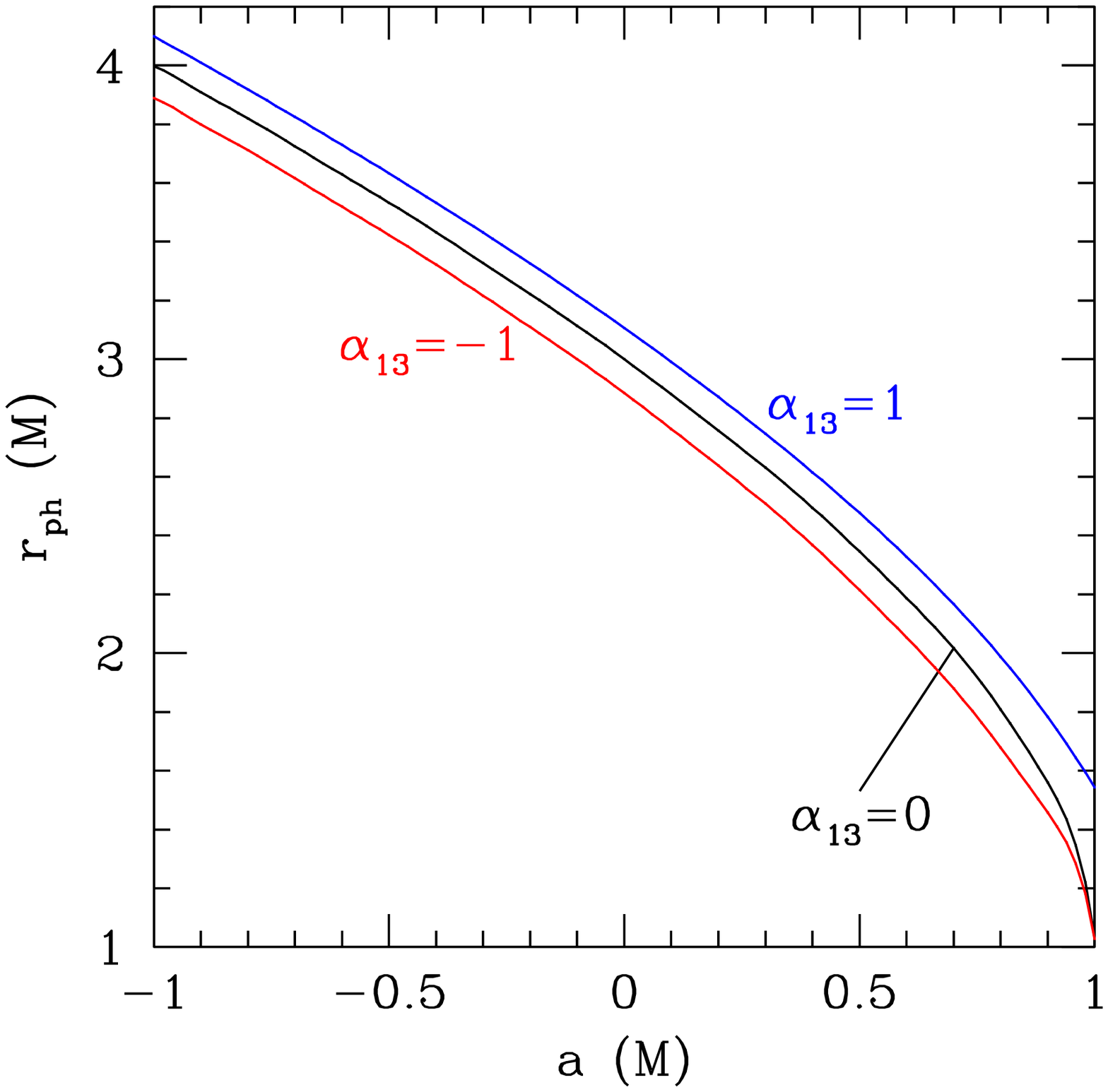,height=3.2in}
\psfig{figure=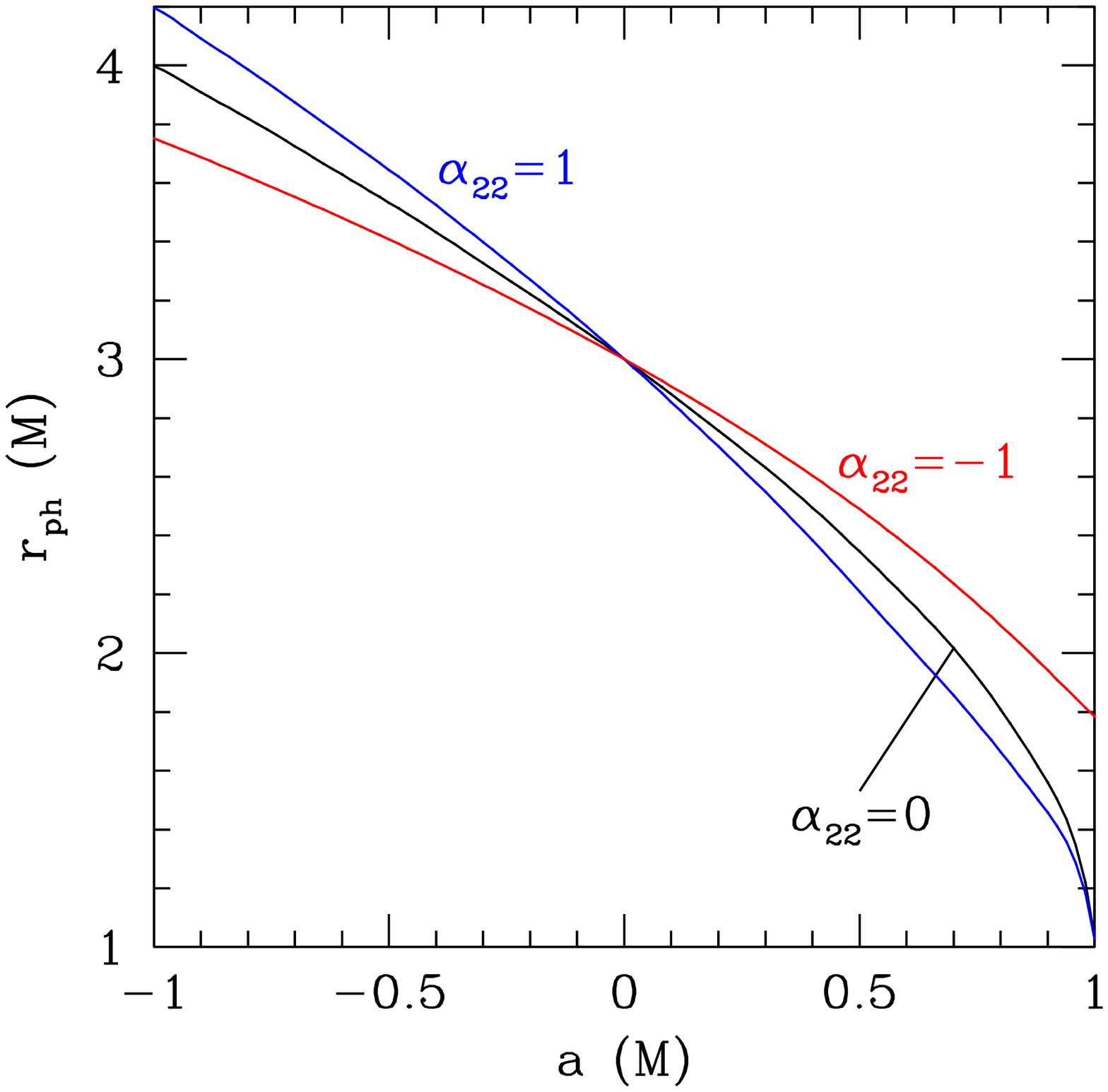,height=3.2in}
\end{center}
\caption{Radius of the circular photon orbit as a function of the spin for different values of the deviation parameters (top) $\alpha_{13}$ and (bottom) $\alpha_{22}$. In each panel, only one deviation parameter is varied, while the other one is set to zero. Top panel: the radius of the circular photon orbit decreases for increasing values of the spin and decreasing values of the deviation parameter $\alpha_{13}$. For negative values of the parameter $\alpha_{13}$, the circular photon orbit merges with the event horizon when $a=M$. Bottom panel: the radius of the circular photon orbit decreases for decreasing values of the parameter $\alpha_{22}$ if $a<0$ and increasing values of the parameter $\alpha_{22}$ if $a>0$. If $a=0$, the circular photon orbit is located at the radius $r_{\rm}=3M$, irrespective of the value of the parameter $\alpha_{22}$.}
\label{fig:photonorbit}
\end{figure}

\begin{figure}[ht]
\begin{center}
\psfig{figure=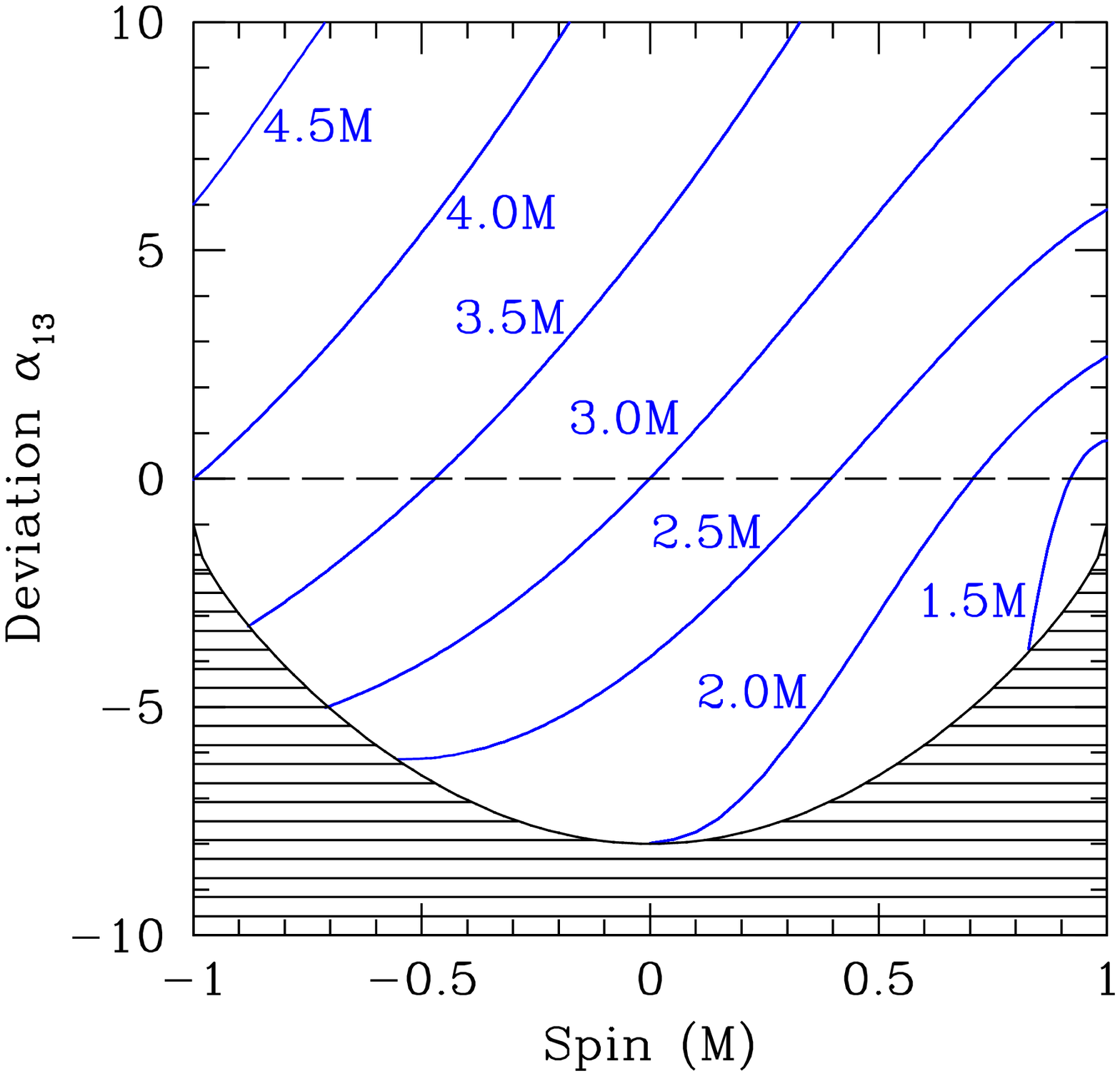,height=3.2in}
\psfig{figure=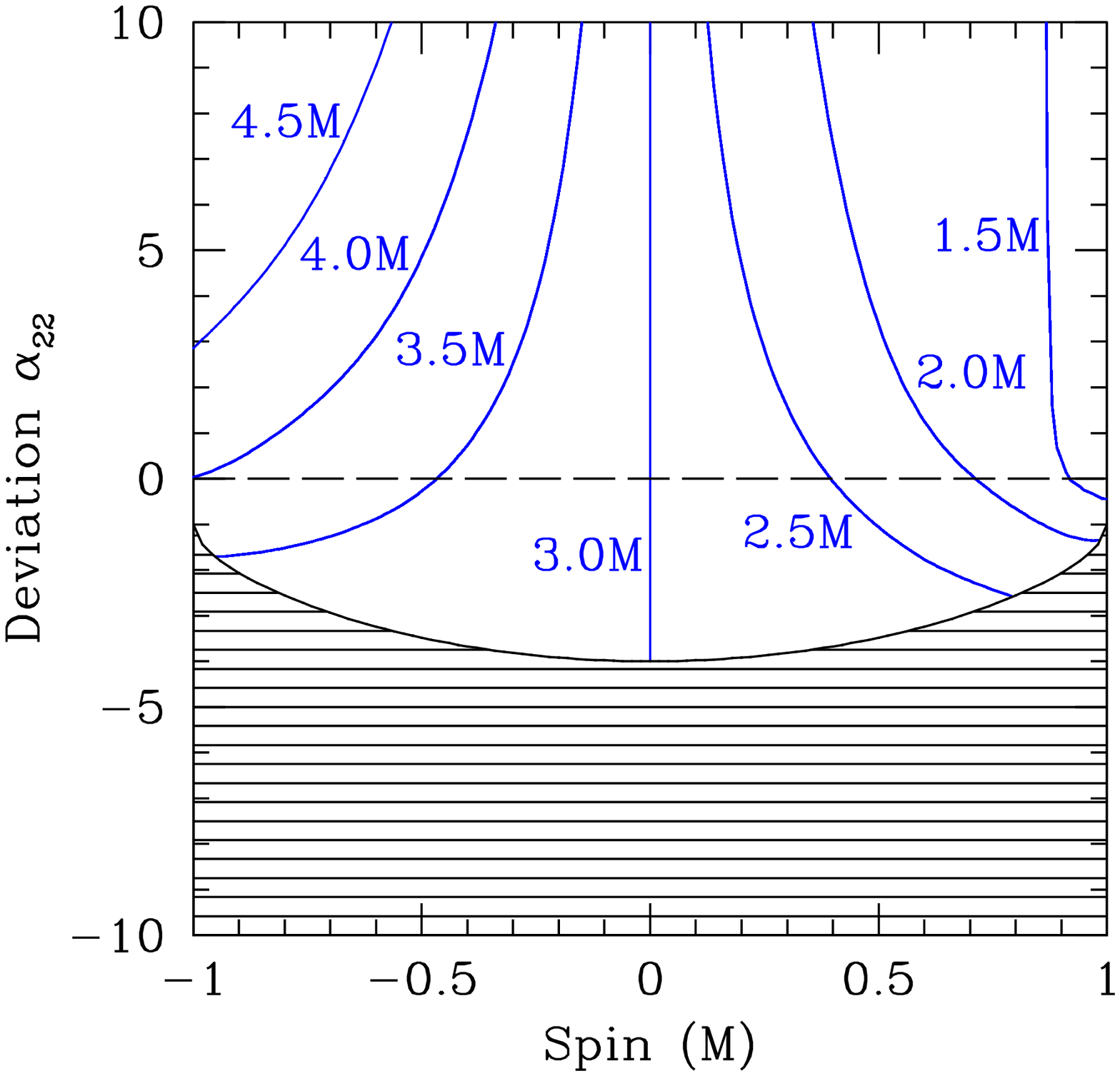,height=3.2in}
\end{center}
\caption{Contours of constant circular photon orbit radius as a function of spin and the deviation parameters (top) $\alpha_{13}$ and (bottom) $\alpha_{22}$. In each panel, only one deviation parameter is varied and the other one is set to zero. In the top panel, the radius of the circular photon orbit decreases for increasing values of the spin and decreasing values of the parameter $\alpha_{13}$. In the bottom panel, the radius of the circular photon orbit decreases for increasing values of the spin and decreasing values of the parameter $\alpha_{22}$ if $a<0$ and increasing values of the parameter $\alpha_{22}$ if $a>0$. If $a=0$, the radius of the circular photon orbit is independent of the parameter $\alpha_{22}$ and is located at a radius of $3M$. The dashed line corresponds to a Kerr black hole. The shaded region marks the excluded part of the parameter space.}
\label{fig:photonorbitcontours}
\end{figure}

In Figure~\ref{fig:photonorbit}, I plot the radius of the circular photon orbit as a function of the spin of the black hole for different values of the parameters $\alpha_{13}$ and $\alpha_{22}$. In Figure~\ref{fig:photonorbitcontours}, I plot contours of constant circular photon orbit radius as a function of the spin and the deviation parameters $\alpha_{13}$ and $\alpha_{22}$. The dependence of the circular photon orbit on these deviation parameters is similar to the dependence of the ISCO on the same deviation parameters (see Johannsen 2013b). The location of the circular photon orbit depends strongly on both deviation parameters $\alpha_{13}$ and $\alpha_{22}$ except for black holes that have a nearly maximal spin $a\sim M$. In that case, the circular photon orbit radius depends only weakly on the deviation parameters. The location of the circular photon orbit decreases for increasing values of the spin and decreasing values of the parameter $\alpha_{13}$. The circular photon orbit radius also decreases for increasing values of the parameter $\alpha_{22}$ if $a>0$ and decreasing values of the parameter $\alpha_{22}$ if $a<0$. For nonspinning black holes, the radius of the circular photon orbit is independent of the parameter $\alpha_{22}$.

In the following, I derive semi-analytic expressions of the images of photon rings around black holes that are described by the metric in Equation~(\ref{eq:metric}). In this derivation, I make use of the existence of three independent constants of motion in this metric.

First I define an inertial reference frame of an observer at a large distance from the black hole (or of any local observer). For the basis vectors, I choose
\ba
e_{\hat t} &=& \zeta e_t + \gamma e_\phi, \nn \\
e_{\hat r} &=& \frac{1}{\sqrt{g_{rr}}} e_r, \nn \\
e_{\hat \theta} &=& \frac{1}{\sqrt{g_{\theta\theta}}} e_\theta, \nn \\
e_{\hat\phi} &=& \frac{1}{\sqrt{g_{\phi\phi}}} e_\phi,
\label{eq:localframe}
\ea
where $\{e_t,e_r,e_\theta,e_\phi\}$ is the coordinate basis of the metric. The basis vectors of the local observer are related to the coordinate basis of the metric via the relations
\be
e_{\hat\alpha} = e_{~\hat\alpha}^\mu e_\mu, 
\ee
where
\be
e_{~\hat\alpha}^\mu e_{~\hat\beta}^\nu g_{\mu\nu} = \eta_{{\hat\alpha}{\hat\beta}}
\label{eq:ONB}
\ee
and where $\eta_{{\hat\alpha}{\hat\beta}}={\rm diag}(-1,1,1,1)$ is the Minkowski metric. The constants $\zeta$ and $\gamma$ in Equation~(\ref{eq:localframe}) have to be chosen in a manner so that the local basis is orthonormal. From the relations in Equation~(\ref{eq:ONB}), I obtain the expressions
\ba
\zeta &=& \sqrt{ \frac{ g_{\phi\phi} }{ g_{t\phi}^2 - g_{tt}g_{\phi\phi} } }, \nn \\
\gamma &=& -\frac{g_{t\phi}}{g_{\phi\phi}} \sqrt{ \frac{ g_{\phi\phi} }{ g_{t\phi}^2 - g_{tt}g_{\phi\phi} } }
\label{eq:zetagamma}
\ea
in terms of the metric elements given by Equation~(\ref{eq:metric}).

In the new basis, the locally measured energy and axial angular momentum are given by the expressions
\ba
p^{\hat t} &=& \zeta E - \gamma L_z, \\
p^{\hat \phi} &=& \frac{1}{\sqrt{g_{\phi\phi}}} L_z.
\ea
I can then define the impact parameters $x'$ and $y'$ in the usual manner (Bardeen 1973), which are the Cartesian coordinate axes of the image plane of the observer located at a distance $r=r_0$ and inclination $\theta=i$:
\ba
x' &\equiv& -r_0\frac{p^{\hat \phi}}{p^{\hat t}} = -r_0 \frac{\xi}{\sqrt{g_{\phi\phi}}\zeta \left( 1 + \frac{g_{t\phi}}{g_{\phi\phi}}\xi \right) }, \\
y' &\equiv& r_0\frac{p^{\hat \theta}}{p^{\hat t}} = r_0 \frac{\pm \sqrt{\Theta(i)} }{\sqrt{g_{\theta\theta}}\zeta \left( 1 + \frac{g_{t\phi}}{g_{\phi\phi}}\xi \right) },
\ea
where (see Johannsen 2013b)
\ba
p^{\hat{\theta}}&=&\frac{p_\theta}{\sqrt{g_{\theta\theta}}}=\frac{\pm\sqrt{\Theta(i)}}{\sqrt{g_{\theta\theta}}}, \\
\Theta(i)&\equiv& \eta + a^2\cos^2 i - \xi^2\cot^2 i.
\ea
In these expressions, all metric elements are evaluated at $r=r_0$ and $\theta= i$ and I introduced the parameters
\ba
\xi &\equiv& \frac{L_z}{E},
\label{eq:xi} \\
\eta &\equiv& \frac{C}{E^2},
\label{eq:eta}
\ea
where $C$ is the generalized Carter constant given in Equation~(\ref{eq:CarterQ}). 

In the limit $r_0\rightarrow\infty$, I obtain the expressions
\ba
x' &=& -\frac{\xi}{\sin i},
\label{eq:xprime} \\
y' &=& \pm \sqrt{\Theta(i)},
\label{eq:yprime}
\ea
which are similar to the ones for the Kerr metric (Bardeen 1973). Note that the $y'$-axis is chosen so that it lies in a plane with the rotation axis of the black hole and that the origin of the image plane is centered on the black hole.

In order to obtain explicit expressions of the parameters $\xi$ and $\eta$ which correspond to the photon ring of the black hole, I determine these parameters from the location of the circular photon orbit. Since these parameters are combinations of the constants of motion, they are conserved along the null geodesics that project the circular photon orbit onto the photon ring in the image plane and can be calculated at any point along these geodesics. This is most easily achieved at the location of the circular photon orbit.

At the circular photon orbit, the $r$-component of the photon four-momentum (Johannsen 2013b),
\be
p^r = \frac{\sqrt{A_5(r)R(r)}}{\tilde{\Sigma}},
\label{eq:prup}
\ee
where
\be
R(r) \equiv [(r^2+a^2)A_1(r) - aA_2(r) \xi]^2 - \Delta [(\xi-a)^2+\eta],
\label{eq:R}
\ee
as well as its radial derivative vanish. Since from Equations~(\ref{eq:A5})--(\ref{eq:lowerbounds}) $\tilde{\Sigma}>0$ and $A_5(r)>0$, I only need to consider the function $R(r)$. Thus, I simultaneously solve the equations
\ba
R(r) &=& 0, \nn \\
\frac{dR(r)}{dr} &=& 0
\label{eq:photonorbit}
\ea
for the parameters $\xi$ and $\eta$. The expressions for these parameters are given explicitly in the Appendix.

\begin{figure}[ht]
\begin{center}
\psfig{figure=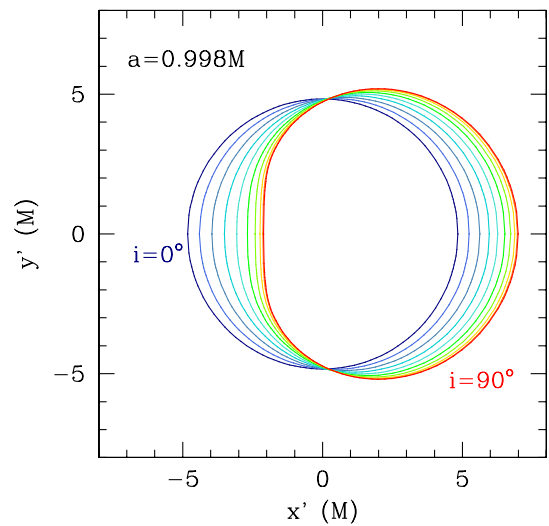,height=3.2in}
\end{center}
\caption{Images of photon rings around a Kerr black hole with spin $a=0.998M$ for different values of the inclination $i$. Photon rings are shown at inclination angles $i=0^\circ$ (dark blue), $i=10^\circ$ (navy blue), ..., $i=90^\circ$ (red). Even for a Kerr black hole with such an extreme spin, the ring shape is nearly circular except for high inclinations.}
\label{fig:kerrspin}
\end{figure}

\begin{figure*}[ht]
\begin{center}
\psfig{figure=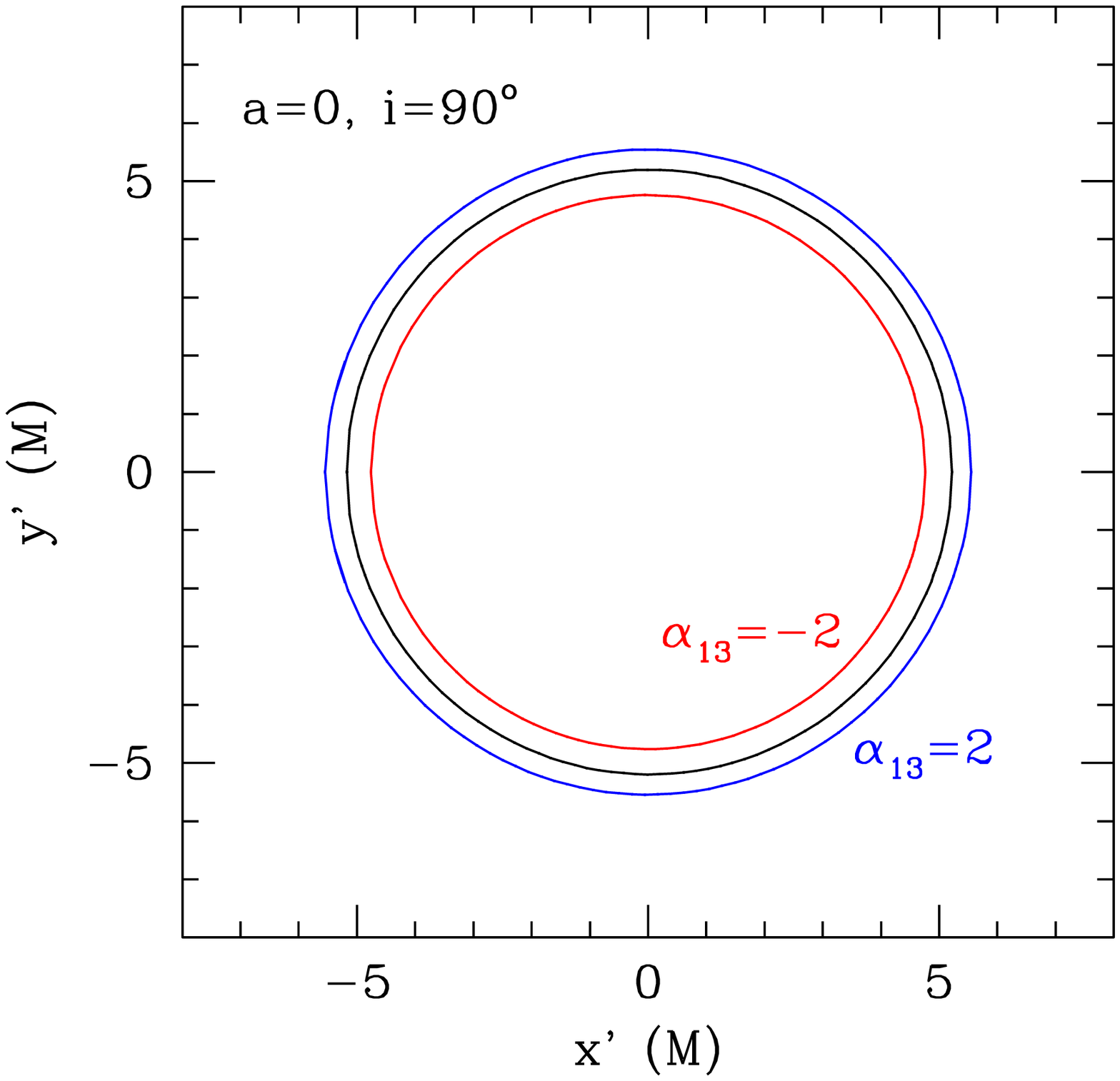,height=2.3in}
\psfig{figure=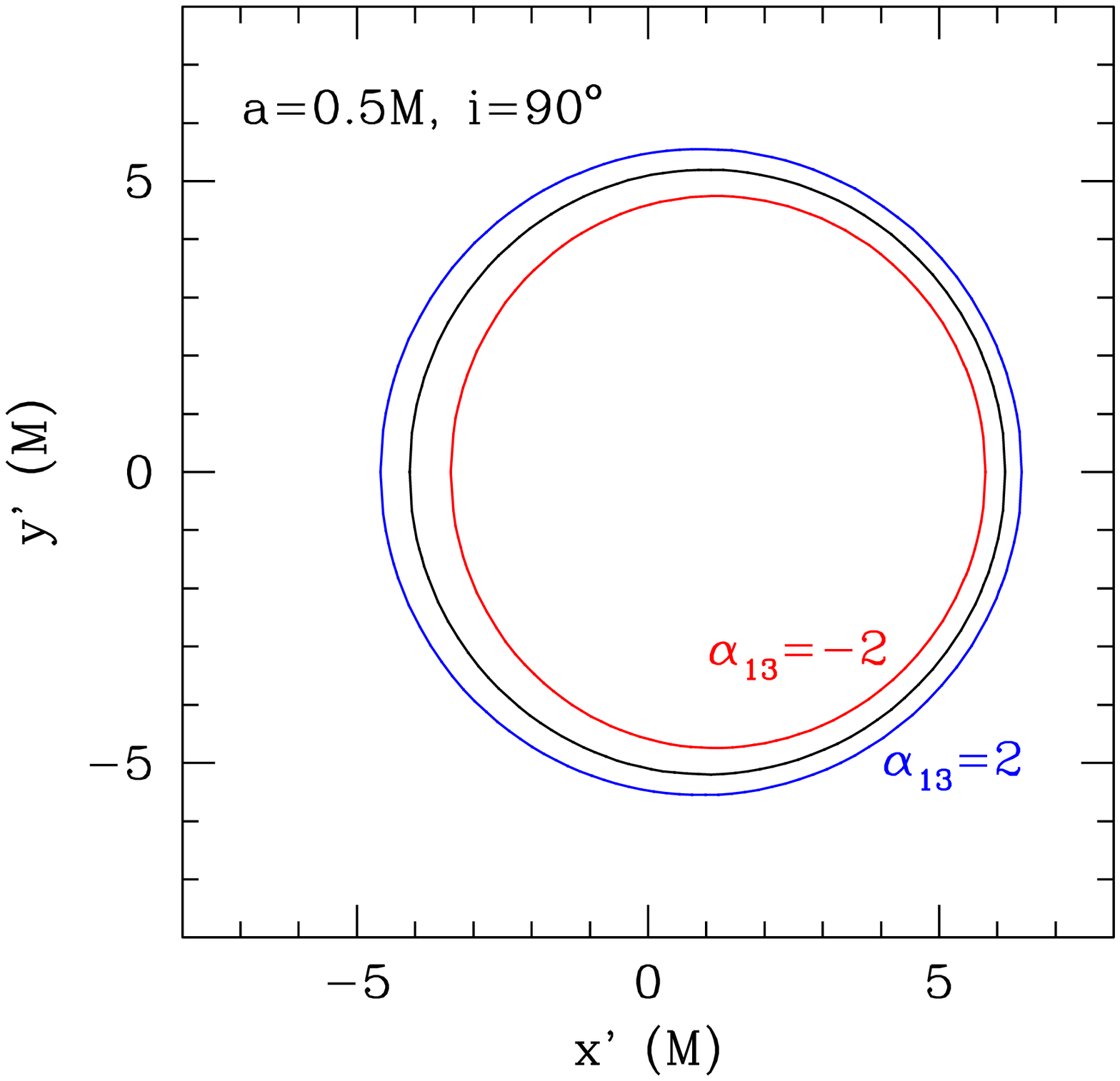,height=2.3in}
\psfig{figure=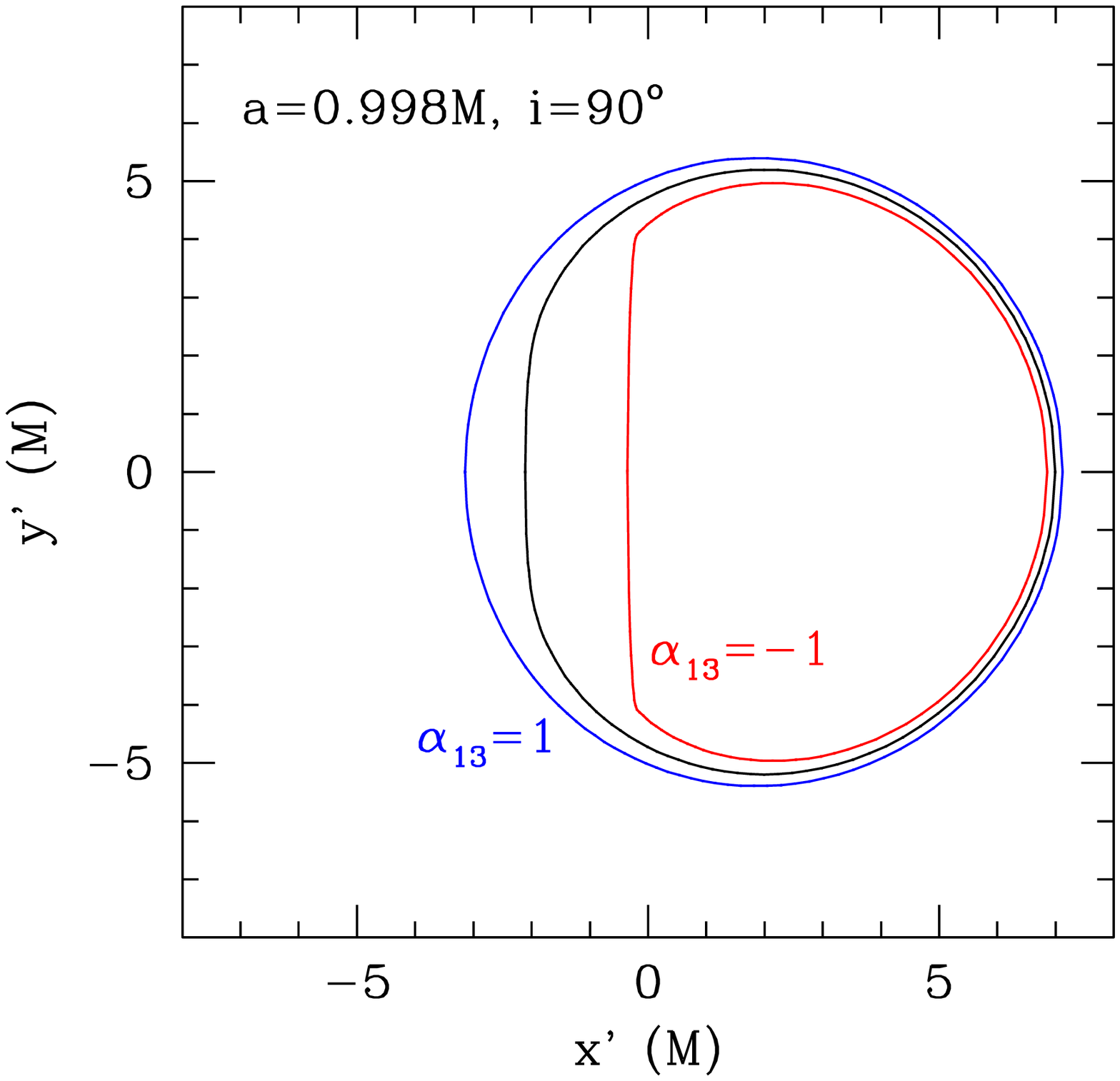,height=2.3in}
\psfig{figure=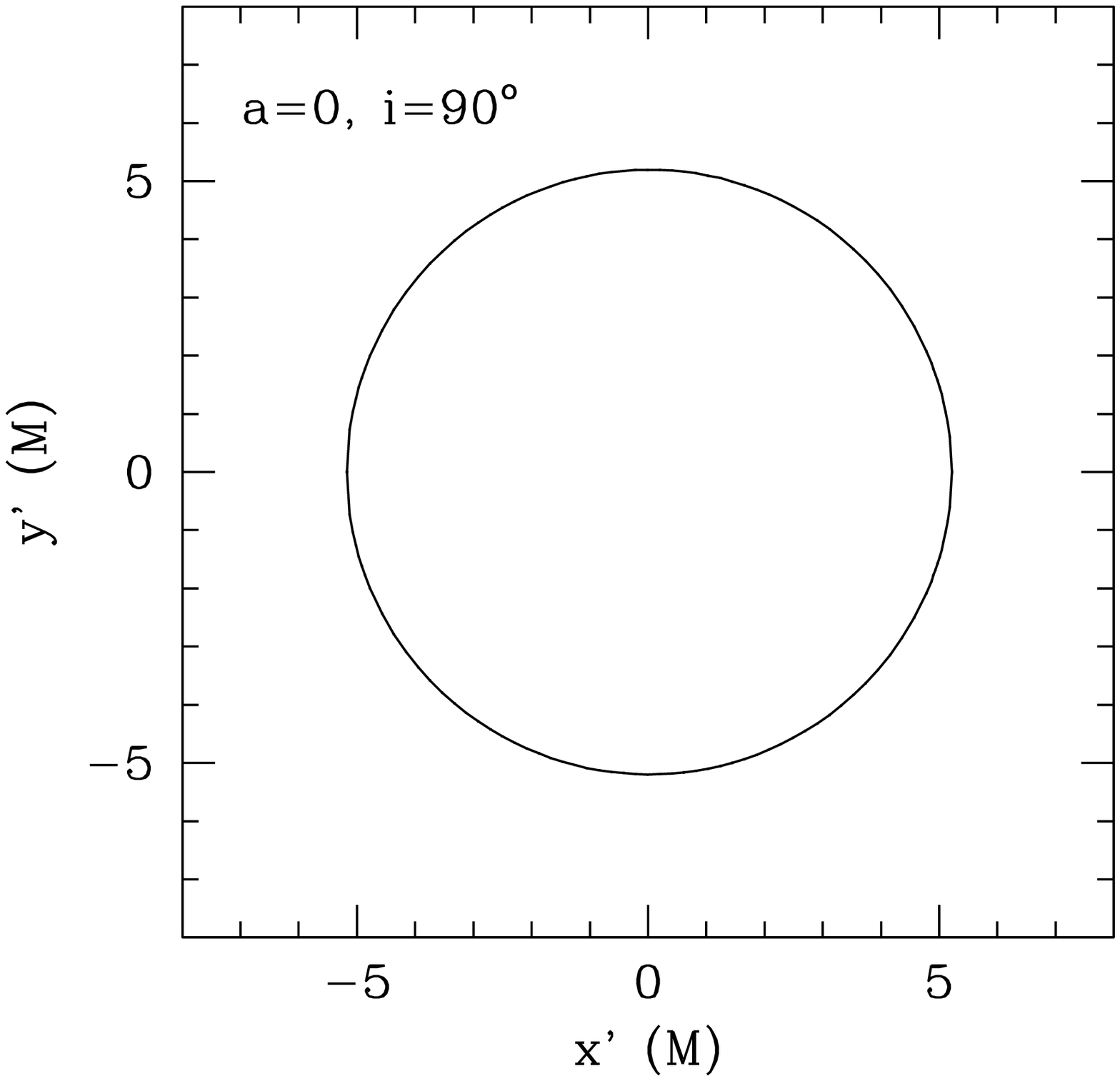,height=2.3in}
\psfig{figure=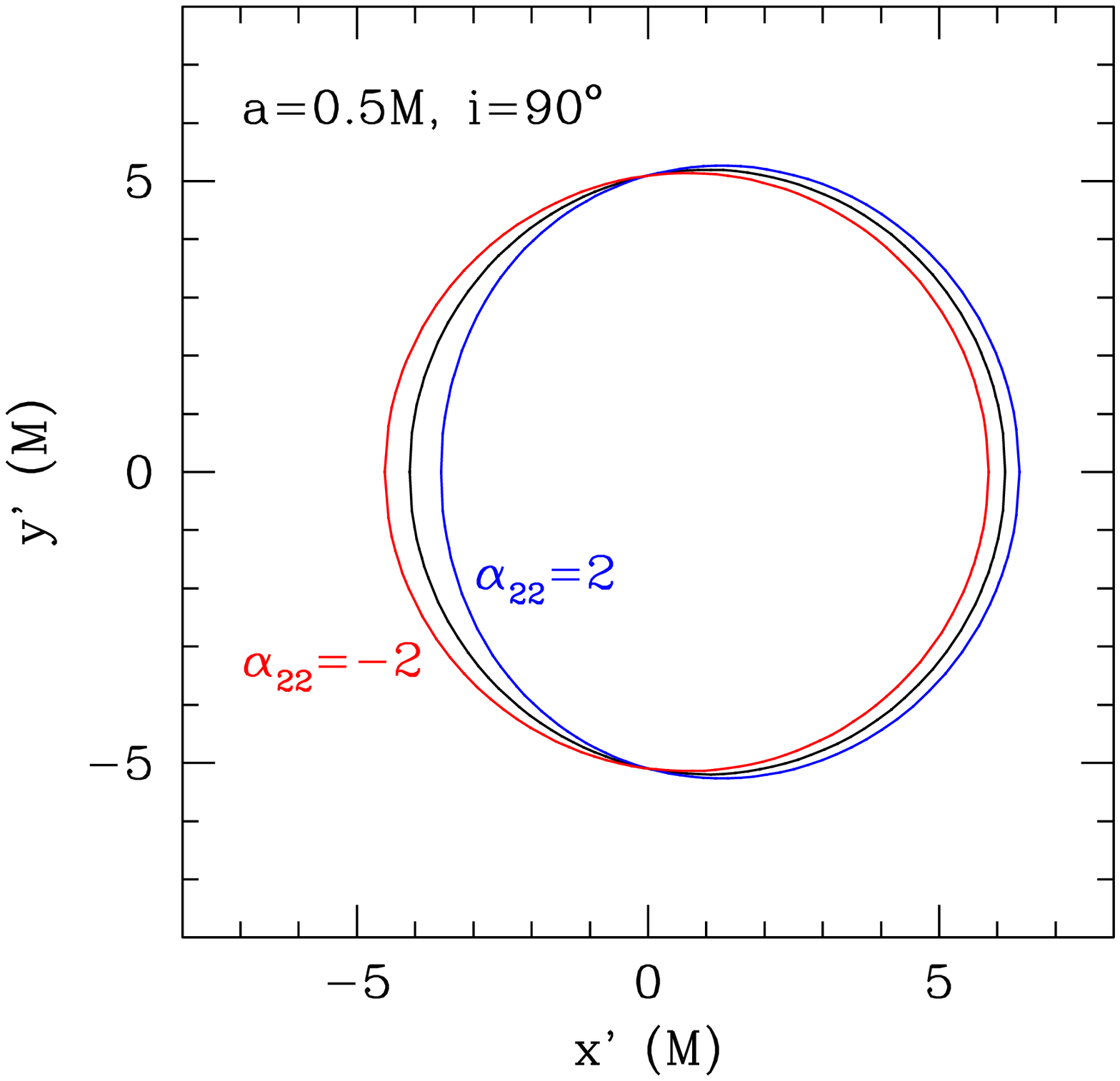,height=2.3in}
\psfig{figure=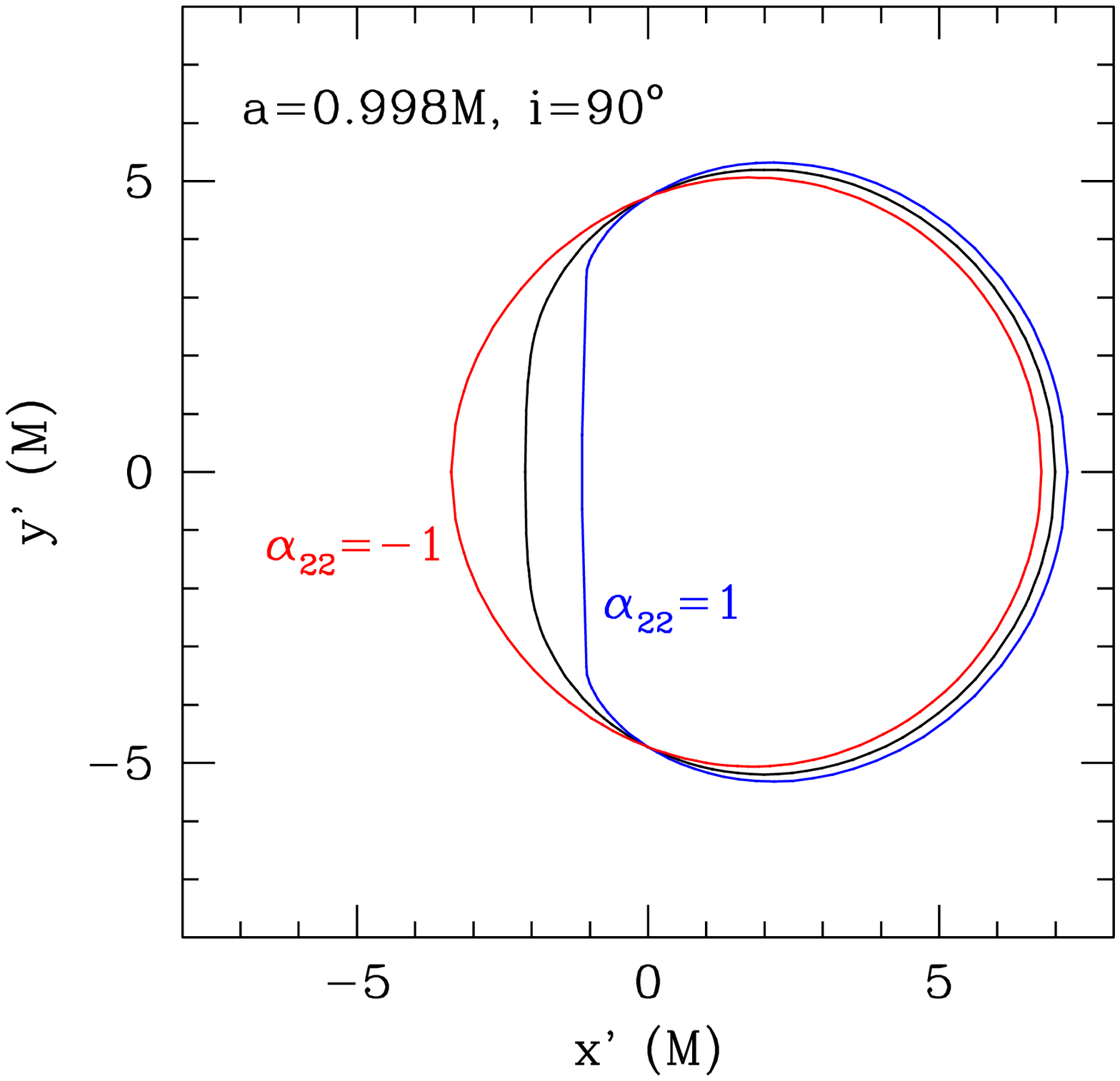,height=2.3in}
\end{center}
\caption{Photon rings around black holes at an inclination angle $i=90^\circ$ for different values of the spin $a$ and deviation parameters $\alpha_{13}$ (top row) and $\alpha_{22}$ (bottom row). The ring size is largely determined by the black hole mass and the parameter $\alpha_{13}$, while it is affected only marginally by the spin and the parameter $\alpha_{22}$. The displacement is primarily affected by the spin of the black hole and depends weakly on the deviation parameters. Nonzero values of the deviation parameters, however, can cause the photon rings to be significantly asymmetric.}
\label{fig:rings}
\end{figure*}

\begin{figure*}[ht]
\begin{center}
\psfig{figure=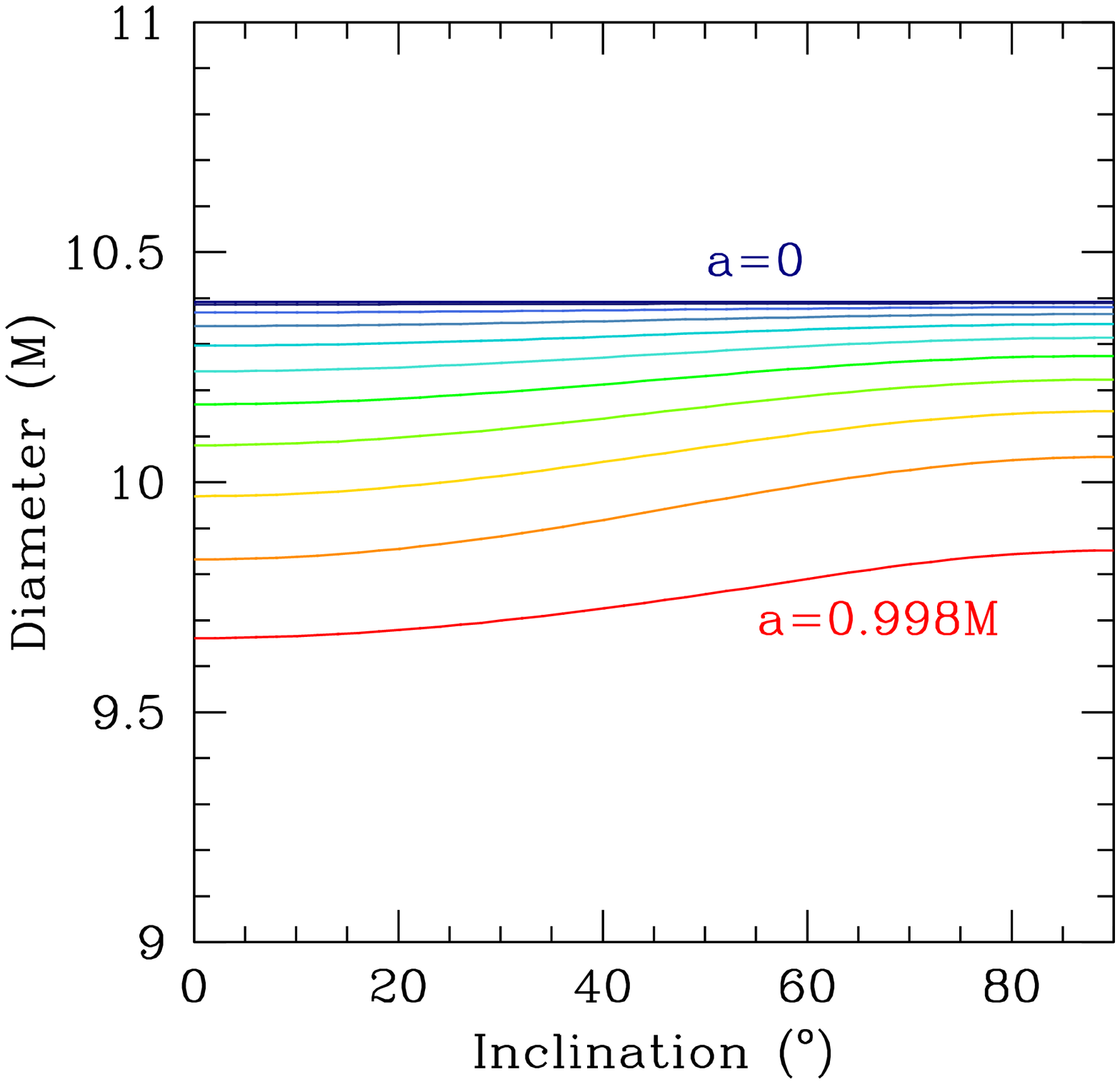,height=2.3in}
\psfig{figure=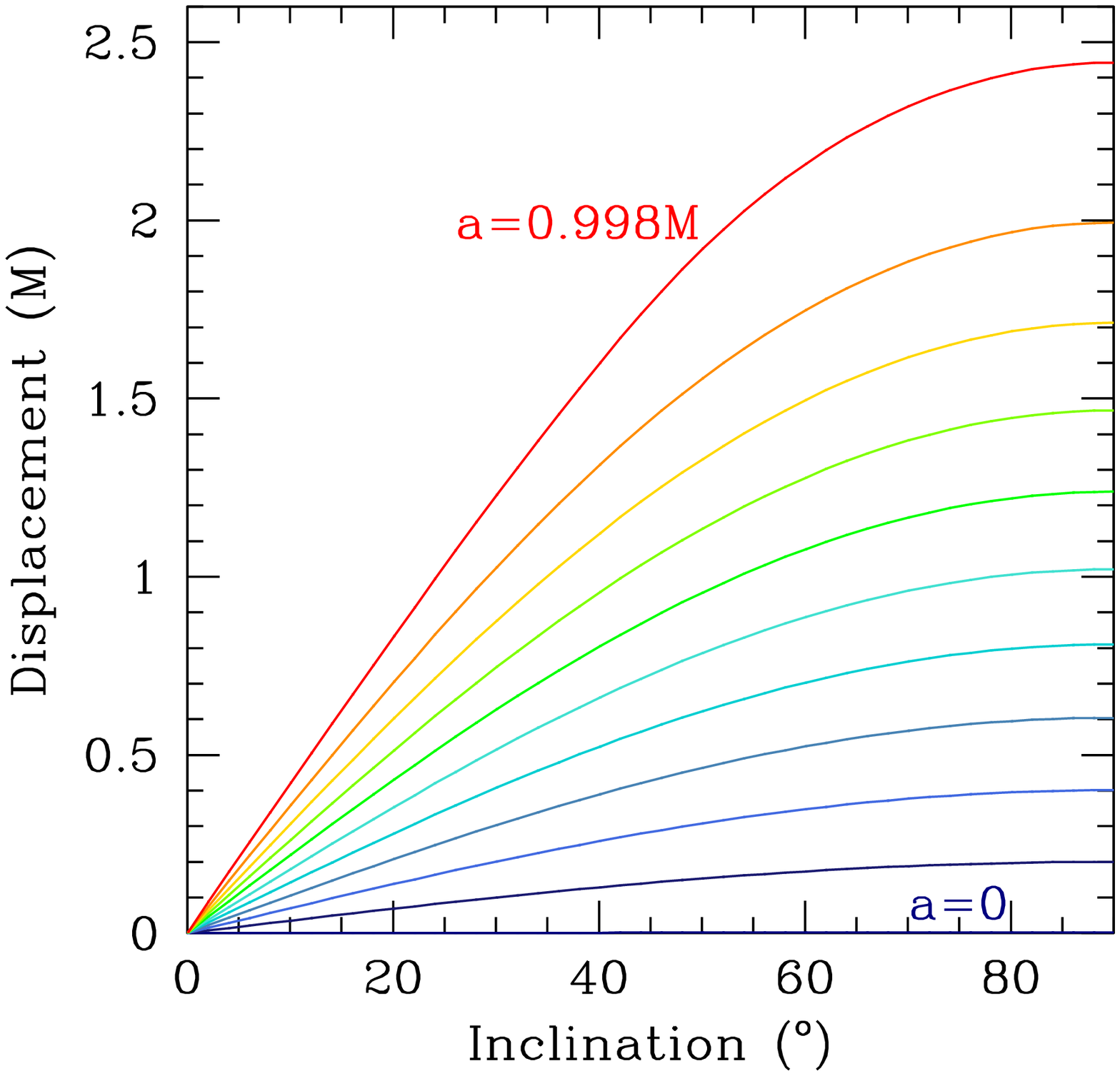,height=2.3in}
\psfig{figure=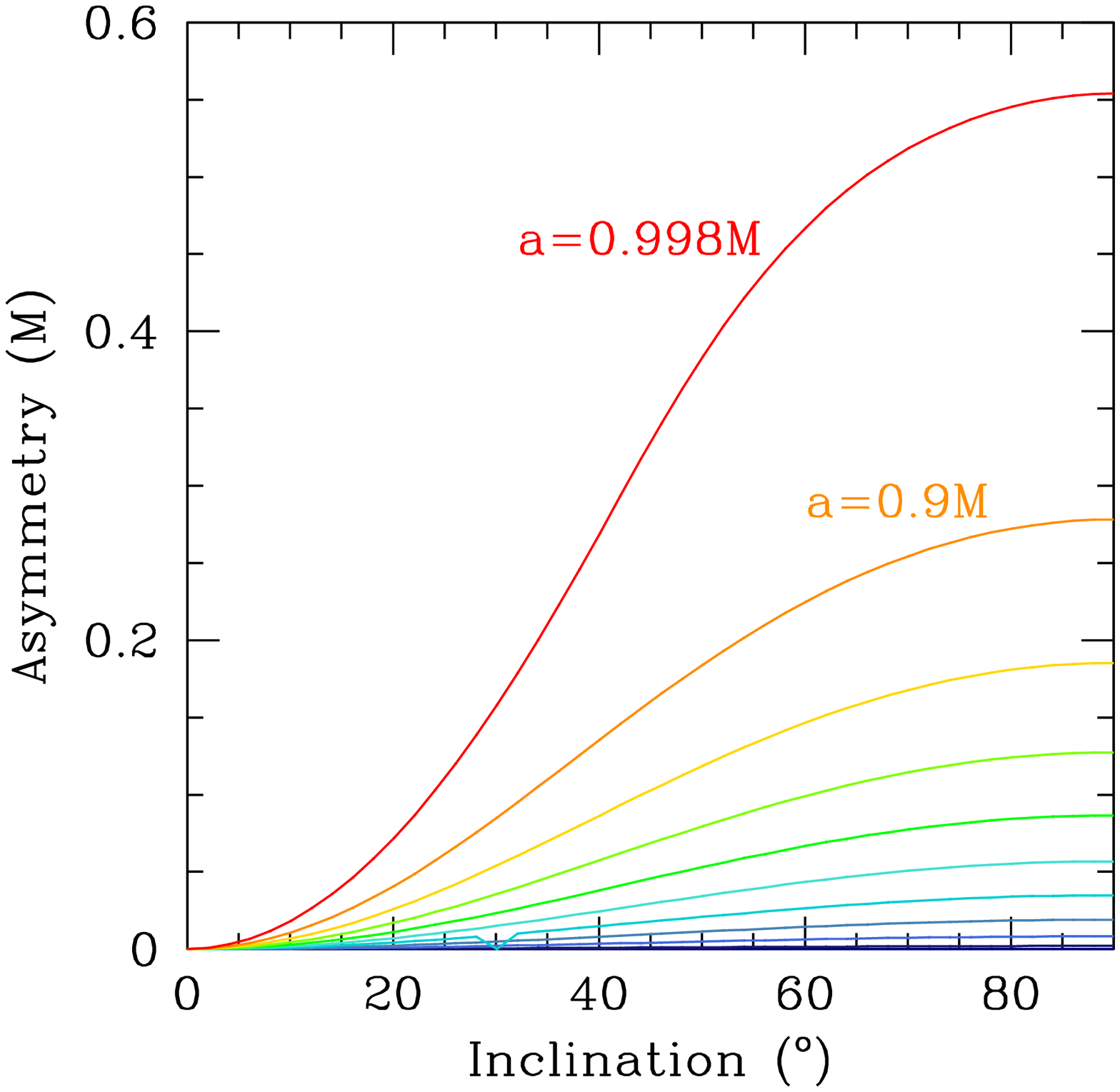,height=2.3in}
\end{center}
\caption{Diameter (left), displacement (center) and asymmetry (right) of the photon ring around a Kerr black hole as a function of the inclination for different values of the spin $a$ in the range $a=0$ (dark blue), $a=0.1M$ (navy blue),..., $a=0.9M$ (orange), $a=0.998M$ (red). The diameter is practically constant at a given value of the spin and depends only very weakly on the observer inclination. The displacement is determined by both the spin and the inclination angle. The asymmetry, however, is small except for very large values of the spin $a\gtrsim0.9M$ at large inclination angles.}
\label{fig:kerrDDA}
\end{figure*}

Images of photon rings around Kerr black holes have already been calculated by several authors (e.g., Bardeen 1973; Luminet 1979; Takahashi 2004; Beckwith \& Done 2005; JP10b; Chan et al. 2013). As an example, I plot in Figure~\ref{fig:kerrspin} images of photon rings around a Kerr black hole with spin $a=0.998M$ for different values of the observer inclination. As discussed in Takahashi (2004), JP10b, and Chan et al. (2013), these images are nearly circular except for very large values of the spin $a\gtrsim0.9M$ and high inclination angles. The image size is primarily determined by the mass of the black hole. For increasing values of the spin at a fixed inclination $i>0^\circ$, the image is displaced off center of the image plane. If the black hole is non-rotating ($a=0$) or viewed face-on ($i=0^\circ$), the photon ring is centered on the black hole. The displacement is reminiscent of the location of the caustics in the Kerr spacetime (Rauch \& Blandford 1994; Bozza 2008).

In Figure~\ref{fig:rings}, I plot images of the photon rings of black holes that are described by the Kerr-like metric in Equation~(\ref{eq:metric}) for different values of the spin and the deviation parameters $\alpha_{13}$ and $\alpha_{22}$ at an inclination angle $i=90^\circ$. The size of the photon ring is altered primarily by the parameter $\alpha_{13}$ and increases for increasing values of this parameter, while the parameter $\alpha_{22}$ affects the image size only marginally. Both deviation parameters only have a weak impact on the image displacement compared to the effect of the spin unless the spin is very large, but they have a strong impact on the shape of the image which can be significantly distorted and asymmetric. However, for positive values of the parameter $\alpha_{13}$ as well as for negative values of the parameter $\alpha_{22}$, photon rings around Kerr-like black holes can be less asymmetric than photon rings around Kerr black holes with the same spin if the spin is very large. Nonetheless, significant ring asymmetries can only be caused by nonzero deviations from the Kerr metric unless the spin is near-extremal ($a\gtrsim0.9M$) and the inclination is large.

\section{Diameter, Displacement, Asymmetry}
\label{sec:DDA}

Following Takahashi (2004) and JP10b, I define the (horizontal) displacement $D$ of the photon ring relative to the center of the image plane by the expression
\be
D \equiv \frac{ \left| x'_{\rm max} + x'_{\rm min} \right| }{ 2 },
\label{eq:displacement}
\ee
where $x'_{\rm max}$ and $x'_{\rm min}$ are the maximum and minimum abscissae of the ring, respectively. Thanks to the reflection symmetry across the equatorial plane of the black hole and the choice of the coordinate system the image plane, there is no displacement in the vertical direction. Following JP10b, I further define the average radius of the ring by the expression
\be
\left< \bar{R} \right> \equiv \frac{1}{2\pi} \int_0^{2\pi} \bar{R} d\alpha,
\ee
where
\ba
\bar{R} &\equiv& \sqrt{ (x'-D)^2+y'^2 }, \\
\tan \alpha &\equiv& \frac{y'}{x'}.
\label{eq:alpha}
\ea
The ring diameter is then
\be
L \equiv 2\left< \bar{R} \right>.
\ee
The ring asymmetry is defined by the expression
\be
A \equiv 2 \sqrt{ \frac{ \int_0^{2\pi} \left(\bar{R} - \left< \bar{R} \right> \right)^2 d\alpha }{ 2\pi } }.
\ee

Takahashi (2004) calculated the ring displacement as a function of the spin for several inclination angles and determined approximately the maximum ring diameter (across the geometric center of the ring) from a set of calculated ring images. JP10b calculated the ring diameter, displacement, and asymmetry for a number of images generated by a ray tracing algorithm in the Kerr and quasi-Kerr spacetimes. They obtained approximate expressions for these ring characteristics in terms of the mass, spin, and quadrupolar deviation parameter $\epsilon$ which was defined in Equation~(\ref{eq:QQK}). Chan et al. (2013) refined the expressions for the diameter and asymmetry of the photon ring around a Kerr black hole using a set of photon ring images simulated by their high-performance ray tracing code {\ttfamily GRay}. In the following, I derive semi-analytic expressions of the ring diameter, displacement, and asymmetry using the expressions of the ring image that I obtained in the previous section. 

Since the ring coordinates $x'$ and $y'$ in the image defined in Equations~(\ref{eq:xprime}) and (\ref{eq:yprime}) are parameterized in terms of the radius $r$, it is convenient to use Equation~(\ref{eq:alpha}) in order to convert the integrals in the ring diameter and asymmetry over the angle $\alpha$ into integrals over the radius by the coordinate transformation
\be
d\alpha = H(r) dr,
\ee
where
\be
H(r) \equiv \left| \frac{d \arctan \frac{y'}{x'} }{dr} \right| = \left| \frac{x' \frac{dy'}{dr} - y' \frac{dx'}{dr} }{ x'^2+y'^2 } \right|
\ee
is the Jacobian of this transformation. The ring diameter and asymmetry then become:
\ba
L &=& \frac{2}{\pi} \int_{r_1}^{r_2} \left< \bar{R} \right> H(r) dr,
\label{eq:diameter} \\
A &=& 2 \sqrt{ \frac{ \int_{r_1}^{r_2} \left(\bar{R} - \left< \bar{R} \right> \right)^2 H(r) dr }{ \pi } },
\label{eq:asymmetry}
\ea
where $r_1$ and $r_2$ are the two roots of the discriminant in the $y'$-coordinate in Equation~(\ref{eq:yprime}) with $r_1<r_2$. Note the extra factor of two in the integration over radius, since the integrals have to be evaluated twice in this radial interval.

In Figure~\ref{fig:kerrDDA}, I plot the diameter, displacement, and asymmetry of the photon ring around a Kerr black hole as a function of the inclination for various values of the spin. Chan et al. (2013) obtained very similar results from fits to a number of simulated photon ring images. For small values of the spin, the diameter is practically constant with a value $\sim10.4M$ (see JP10b). For values of the spin $0\leq|a|\lesssim 0.77M$, the ring diameter lies in the range $10M\lesssim L \lesssim 10.4M$. For large values of the spin $a\gtrsim0.77M$, the ring diameter can be smaller than $10M$. The displacement depends approximately linearly on the spin and the sine of the inclination angle (JP10b; Chan et al. 2013; c.f., Bozza 2008). The asymmetry is generally small and only significant at very large values of the spin $a\gtrsim0.9M$ and high inclinations.

\begin{figure*}[ht]
\begin{center}
\psfig{figure=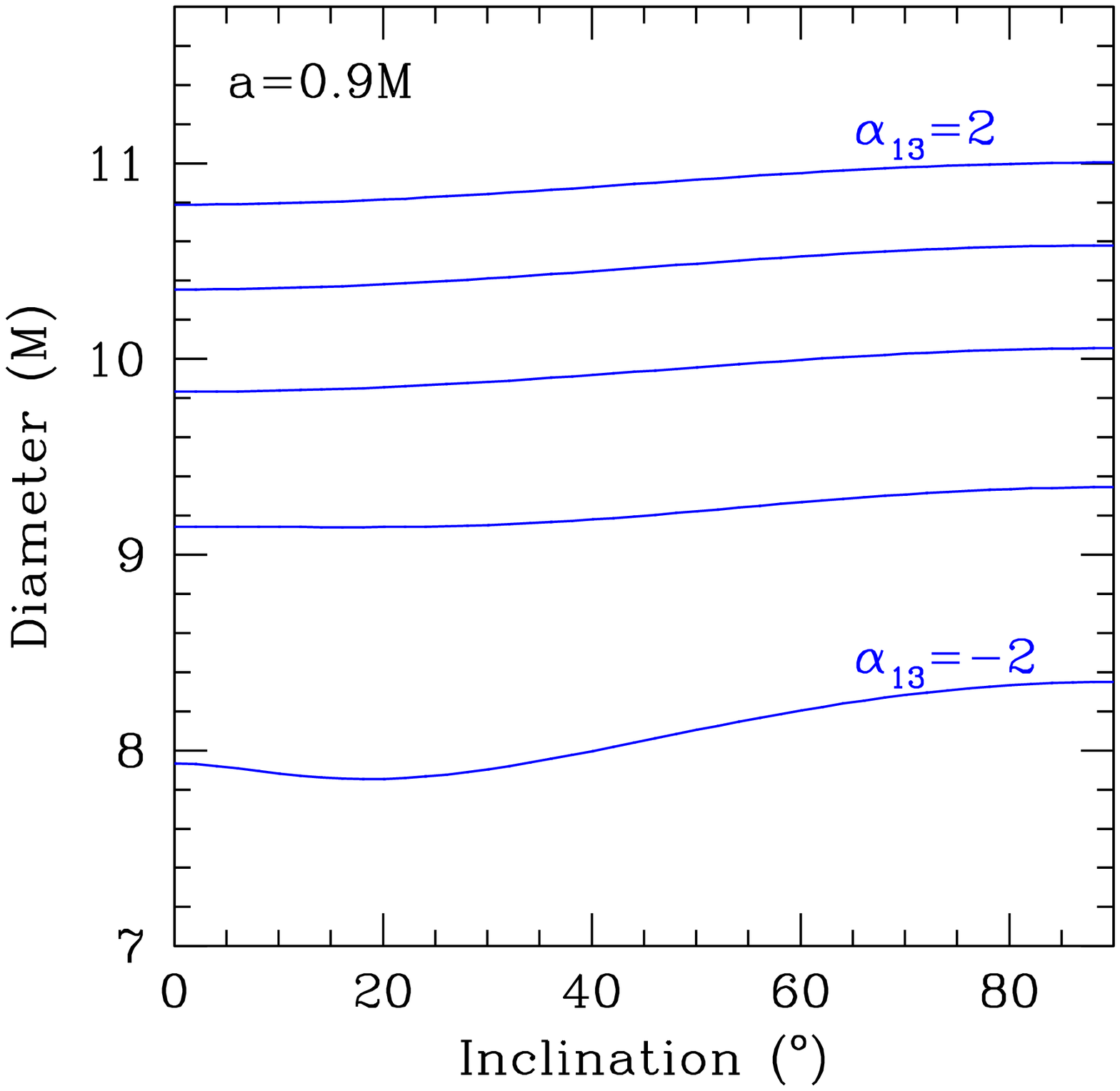,height=2.3in}
\psfig{figure=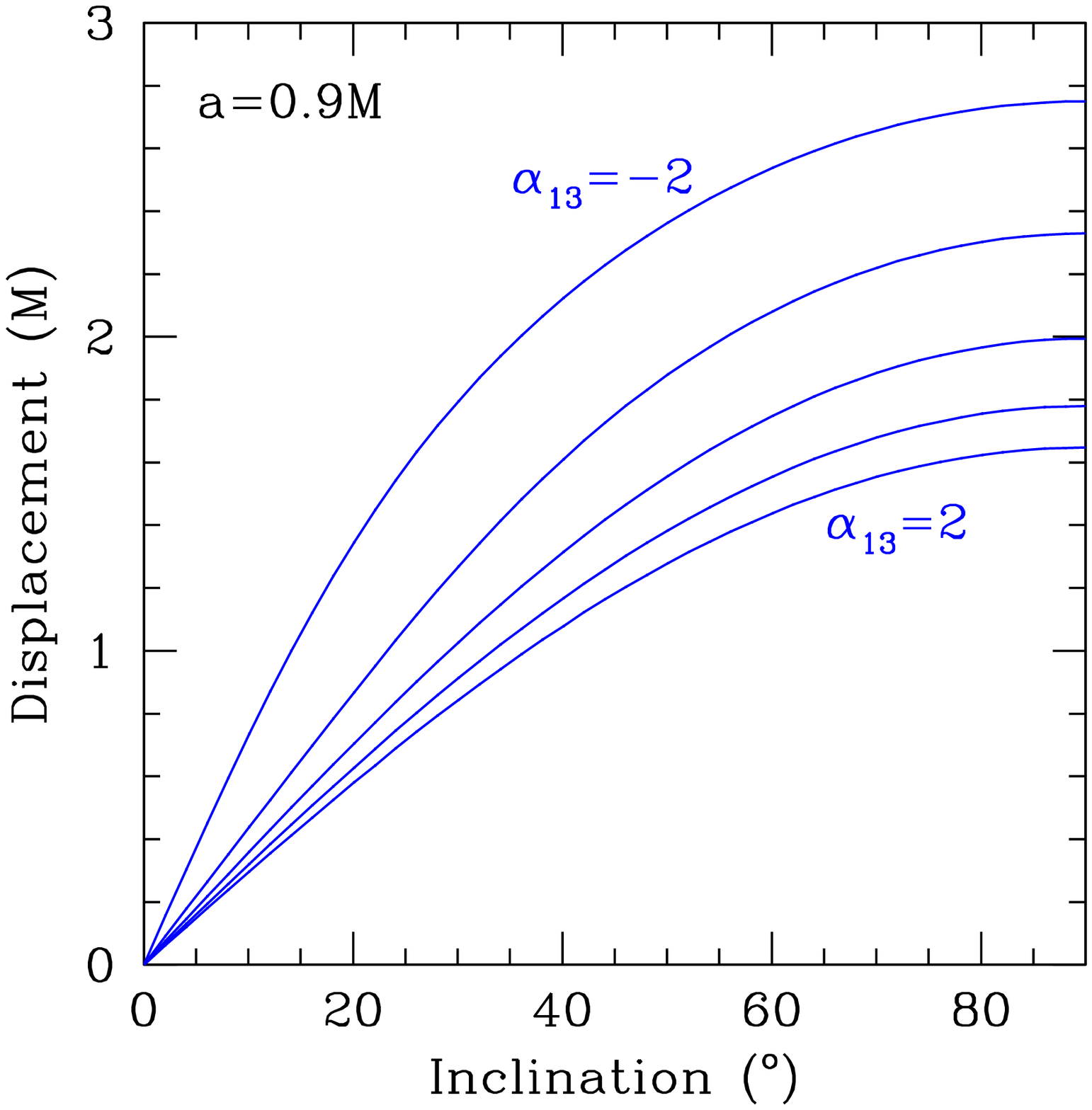,height=2.3in}
\psfig{figure=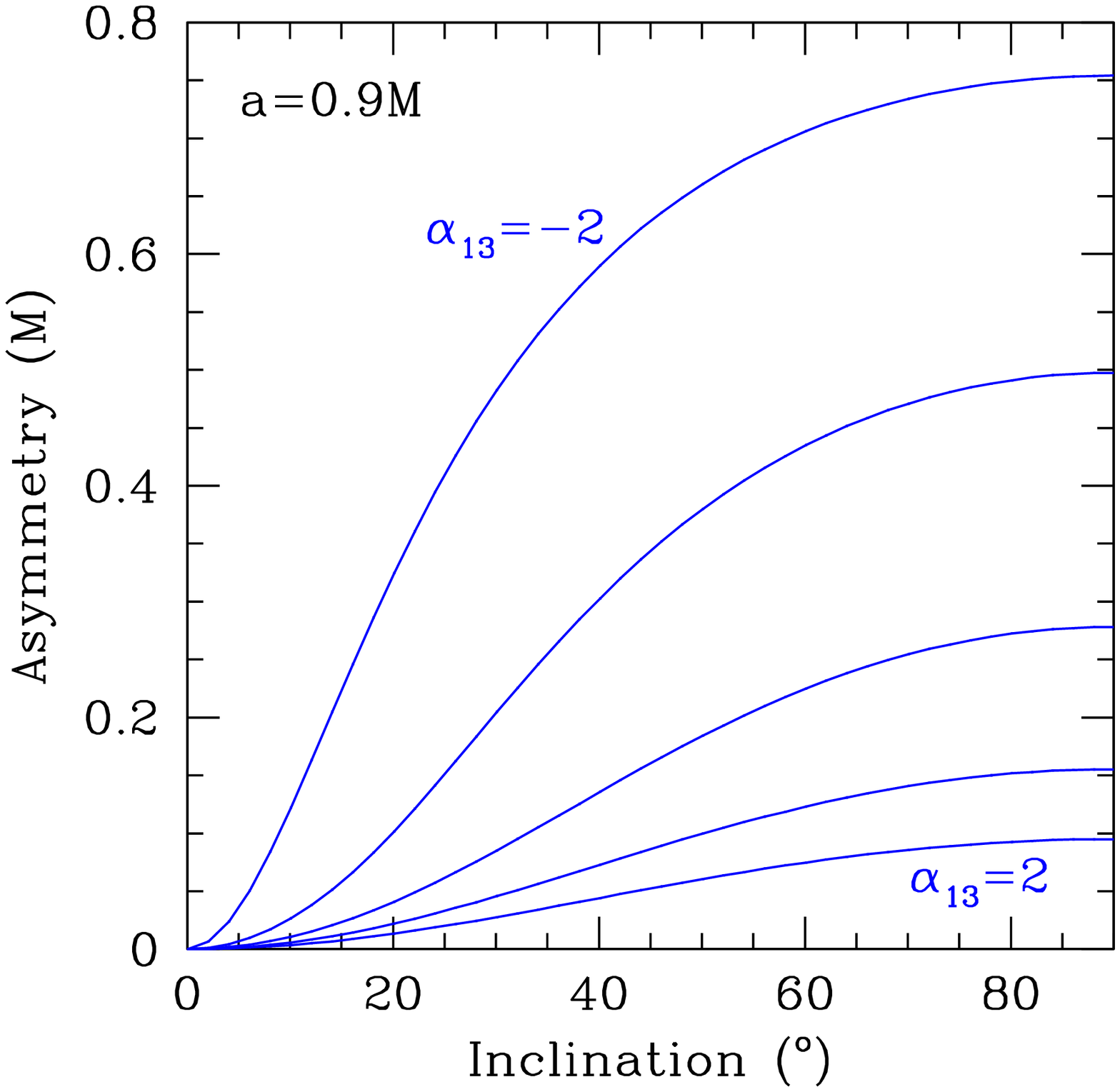,height=2.3in}
\psfig{figure=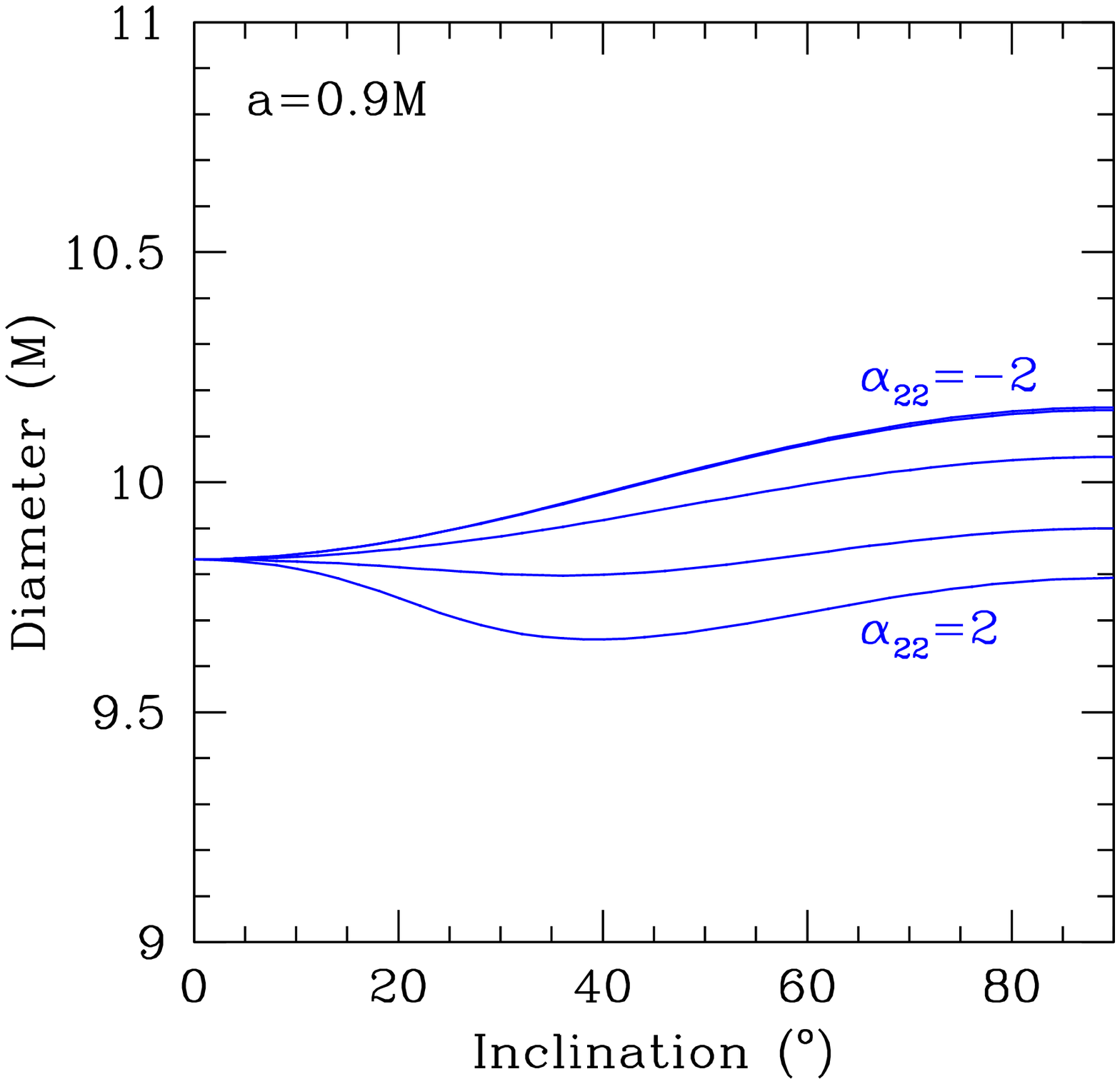,height=2.3in}
\psfig{figure=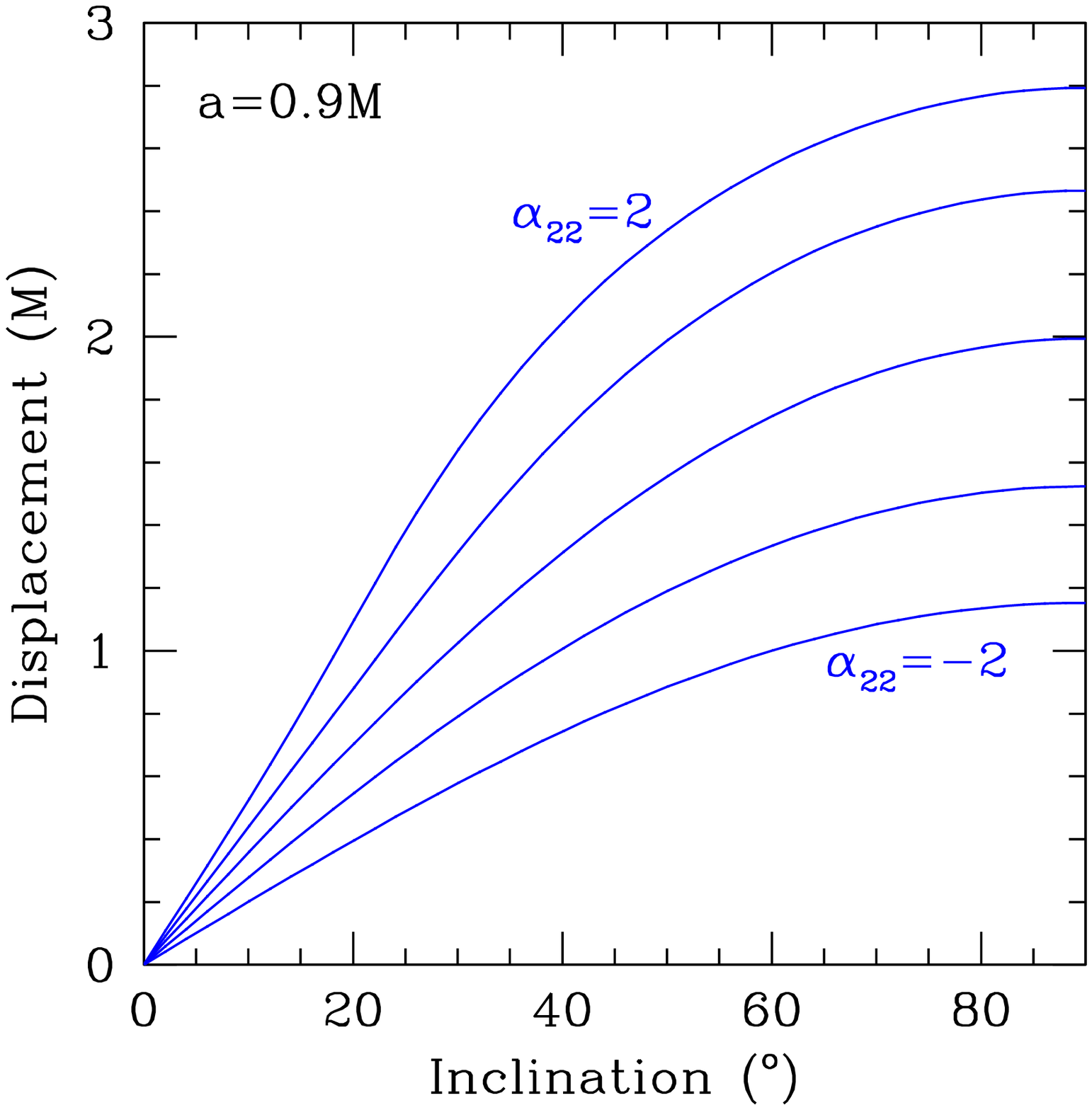,height=2.3in}
\psfig{figure=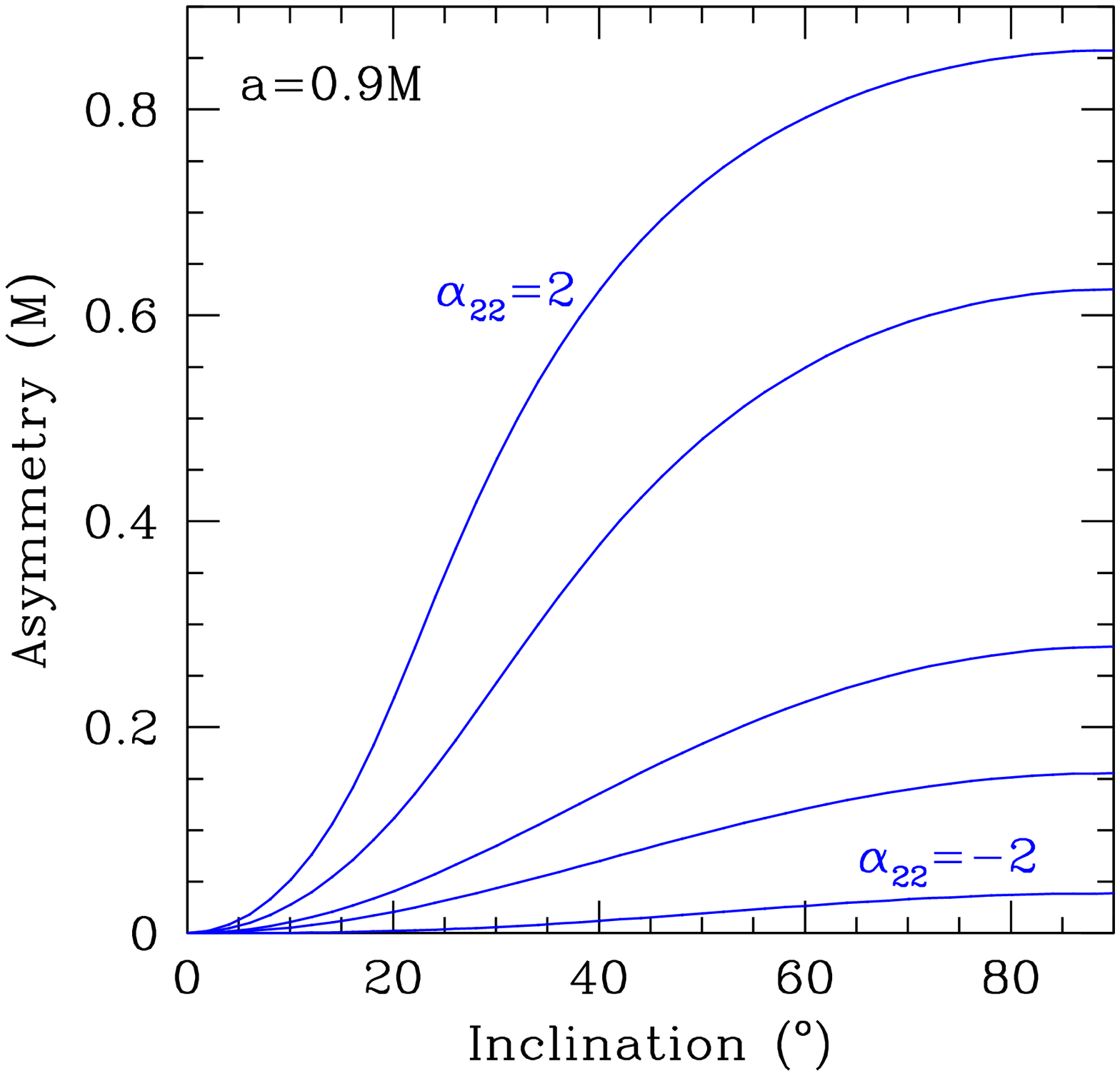,height=2.3in}
\end{center}
\caption{Diameter (left), displacement (center), and asymmetry (right) of photon rings around Kerr-like black holes with a spin $a=0.9M$ as a function of the inclination angle for different values of the deviation parameters $\alpha_{13}$ (top row) and $\alpha_{22}$ (bottom row). The ring diameter depends only weakly on the parameter $\alpha_{22}$, while it is practically constant for fixed values of the parameter $\alpha_{13}$. The deviation parameter $\alpha_{13}$ can be inferred from a measurement of the ring diameter if the mass and distance to the black hole are known. The displacement of the ring is altered by the parameters $\alpha_{13}$ and $\alpha_{22}$ but depends primarily on the spin. Negative values of the parameter $\alpha_{13}$ and positive values of the parameter $\alpha_{22}$ can cause the ring shape to be significantly more asymmetric than the photon ring around a Kerr black hole with the same spin. The ring asymmetry is a direct measure of the degree to which the no-hair theorem is violated.}
\label{fig:rings09}
\end{figure*}

In Figure~\ref{fig:rings09}, I plot the diameter, displacement, and asymmetry of the photon ring around a black hole described by the metric in Equation~(\ref{eq:metric}) with a spin $a=0.9M$ as a function of the inclination for different values of the deviation parameters $\alpha_{13}$ and $\alpha_{22}$. The ring diameter depends only weakly on the parameter $\alpha_{22}$ and is practically constant for fixed values of the parameter $\alpha_{13}$ ranging from $L\approx8M$ for $\alpha_{13}=-2$ to $L\approx10.8M$ for $\alpha_{13}=2$. Therefore, the value of the deviation parameter $\alpha_{13}$ could be inferred from a measurement of the ring diameter if the mass and distance to the black hole are known with sufficient accuracy. The ring displacement, on the other hand, is determined primarily by the spin of the black hole. Even at an inclination angle of $i=90^\circ$, values of the deviation parameters $\alpha_{13}=\pm2$ or $\alpha_{22}=\pm2$ change the displacement caused by the spin alone by at most $50\%$. While the spin could, in principle, be inferred from the ring displacement, it is unlikely to yield a useful spin measurement in practice, because the exact location of the black hole in the image would have to be determined. The asymmetry of the ring is very large if $\alpha_{13}\ll0$ or if $\alpha_{22}\gg0$. For values of the deviation parameters $\alpha_{13}=-2$ and $\alpha_{22}=2$, respectively, the asymmetry is $\approx2.7$ and $\approx3$ times as large as the asymmetry of the ring around a Kerr black hole with the same spin at an inclination of $i=90^\circ$. As in the case of the quasi-Kerr metric (see JP10b), the ring asymmetry is, therefore, a direct measure of the degree to which the no-hair theorem is violated.

The modified displacement and asymmetry of photon rings around Kerr-like black holes correspond to the shifted location of the circular photon orbit as shown in Figure~\ref{fig:photonorbit}. For negative values of the parameter $\alpha_{13}$ as well as for positive values of the parameter $\alpha_{22}$, the radius of the circular photon orbit decreases if the black hole spin lies in the range $0<a<M$. Therefore, the photons that comprise the image of the circular photon orbit will experience stronger lightbending which causes the image to become more displaced and more asymmetric as in the case of a the ring around a Kerr black hole with the same spin. Likewise, at least for deviations that are measured in terms of the parameters $\alpha_{13}$, the ring diameter decreases. 

Next, I obtain approximate expressions of the diameters, displacements, and asymmetries from a large number of simulated photon rings. First, I obtain fit formulae for these properties of photon ring around Kerr black holes. I computed the diameters, displacements, and asymmetries of 17,113 photon rings from Equations (\ref{eq:diameter}), (\ref{eq:displacement}), and (\ref{eq:asymmetry}) choosing spin values in the range from $a=0.01M$ to $a=0.99M$ in steps of $\Delta a=0.01M$ and from $a=0.991M$ to $a=0.999M$ in steps of $\Delta a=0.001M$ for values of the inclination ranging from $i=0.01$ to $i=1.57$ in steps of $\Delta i=0.01$.

I make the following ansatz for the diameter, displacement, and asymmetry of the photon ring, where the sub- and superscripts ``K'' refer to a Kerr black hole:
\ba
L_{\rm K}(i,a) &\approx& L_1^{\rm K} + L_2^{\rm K}\cos\left(n_1^{\rm K} i+\varphi^{\rm K}\right),
\label{eq:L_Kerrfit} \\
D_{\rm K}(i,a) &\approx& D_1^{\rm K} \sin\left(n_2^{\rm K} i\right),
\label{eq:D_Kerrfit} \\
A_{\rm K}(i,a) &\approx& \frac{1}{A_1^{\rm K} i^{n_3^{\rm K}} + A_2^{\rm K} i^{n_4^{\rm K}}}.
\label{eq:A_Kerrfit}
\ea
Using this ansatz, I perform a least-squares fit and obtain the expressions
\ba
L_1^{\rm K} &=& 10.3907 - 0.520134a^{2.19644}-0.109423a^{11.5509}, \nn \\
L_2^{\rm K} &=& 0.141644a^{2.06874}+0.077576a^{2.80628} \nn \\
&& -0.305764a^{4.44562}+0.603316a^{6.58314} \nn \\
&& -0.801615a^{9.97291} +0.892182a^{15.1866} \nn \\
&& -0.508699a^{18.8103}, \nn \\
n_1^{\rm K} &=& -2.00026-0.0424525a^{6.43112} \nn \\
&& +0.0503142a^{36.5515}+0.118102a^{153.450}, \nn \\
\varphi^{\rm K} &=& 3.14246+0.0970934a^{5.40872} \nn \\
&& +0.0964881a^{13.3813}-0.0990121a^{196.868}, \nn \\
D_1^{\rm K} &=& 2a + 0.188324a^{3.29029}+0.153158a^{10.4479} \nn \\
&& +0.0982701a^{48.1825}+0.0500847a^{328.089}, \nn \\
n_2^{\rm K} &=& 1-0.0149883a^{0.656102}+0.0188647a^{0.729558} \nn \\
&& +0.0370128a^{2.49086}+0.147758a^{30.2625} \nn \\
&& -0.163190a^{31.1614} -0.0148721a^{259.142}, \nn \\
A_1^{\rm K} &=& -\frac{3.29826}{(1-a)^{0.0708700}}+\frac{4.06676}{a^{1.96447}}+\frac{1.69824}{a^{2.06984}} \nn \\
&& +0.535821a^{58.9636}+0.801151a^{495.375}, \nn \\
n_3^{\rm K} &=& -2.55018+0.679338(1-a)^{0.306884} \nn \\
&& +0.263229a^{1.12755}+0.113085a^{2.87034} \nn \\
&& -0.0657765a^{19.0048} -0.0497249a^{60.8532}, \nn \\
A_2^{\rm K} &=& \frac{0.0482563}{(1-a)^{0.291648}}+\frac{1.17042}{a^{1.99594}}-0.259672a^{7.55857} \nn \\
&& -0.306290a^{55.7992}-0.323404a^{475.326}, \nn \\
n_4^{\rm K} &=& 0.235657+1.39539(1-a)^{0.755126} \nn \\
&& +1.33468a^{1.12230}-0.842482a^{1.85033} \nn \\
&& +0.0962717a^{30.9888} +0.140634a^{119.303} \nn \\
&& +0.127251a^{506.431}.
\ea

In the fit of the displacement, I held the first terms in the parameters $D_1$ and $n_2$ fixed so that $D_{\rm K}(i,a)\approx2a\sin i$ as found in JP10b. The fit formula of the ring diameter is accurate to $<0.09\%$ for spin values $a\leq0.997M$ and to $<0.16\%$ for spins $a\leq0.999M$. The fit formula of the displacement is accurate to $<1.1\%$ for spin values $a\leq0.97M$, to $<2.4\%$ for spins $a\leq0.98M$, and to $<3.7\%$ for spins $a\leq0.999M$. For values of the asymmetry $A\geq0.01M$, the asymmetry fit is accurate to $<1\%$ for spin values $a\leq0.6M$, to $<4.8\%$ for spins $a\leq0.98M$, and to $<19.4\%$ for spins $a\leq0.999M$. The largest uncertainties in the fit occur at low inclinations, and the fit is accurate at all spin values to $<2.5\%$ for inclinations $i\geq22.4^\circ$. Values of the ring asymmetry $A<0.01M$, which occur only at spin values $a\sim0$ and inclinations $i\sim0$, were partly affected by numerical uncertainty and, therefore, neglected in the fit. Since these values of the asymmetry are very small, this affects the fit only marginally. The asymmetry can also be fitted with an ansatz of the form $A_0\sin^n i$ as in JP10b and Chan et al. (2013). Such a fit, however, introduces comparatively large errors at high spins, where the asymmetry deviates significantly from a sinusoidal form.

Next, I obtain approximate expressions for the diameters, displacements, and asymmetries of photon rings around the Kerr-like black holes described by Equation~(\ref{eq:metric}). I calculated the diameters, displacements, and asymmetries of 288,300 photon rings with spins $a/M=0.05,~0.10, ...,~0.95,~0.998$, inclinations $i=0.1,~0.2, ...,~1.5$, and deviations $\alpha_{13}=-1,~-0.9, ...,~2$, $\alpha_{22}=-1,~-0.9, ...,~2$. For the respective fit functions, I introduced polynomial functions of the deviations parameters $\alpha_{13}$ and $\alpha_{22}$ to the corresponding fits for Kerr photon rings in Equations~(\ref{eq:L_Kerrfit})--(\ref{eq:A_Kerrfit}) given by the expressions
\ba
L(i,a,\alpha_{13},\alpha_{22}) &\approx& L_1^{\rm K}+L_1\alpha_{13}+L_2\alpha_{13}^2 +L_3\alpha_{22} \nn \\
&& +L_4\alpha_{22}^2 +L_2^{\rm K}(1+L_5\alpha_{13}) \nn \\
&& \times (1+L_6\alpha_{22}) \cos\left(n_1^{\rm K}i+\varphi^{\rm K}\right),
\label{eq:L_fit1} \\
D(i,a,\alpha_{13},\alpha_{22}) &\approx& D_1^{\rm K} \left(1+D_1 \alpha_{13}+D_2\alpha_{13}^2\right) \nn \\
&& \times \left(1+D_3\alpha_{22}+D_4\alpha_{22}^2\right) \nn \\
&& \times \sin\left[ n_2^{\rm K}(1+n_1\alpha_{13}+n_2\alpha_{13}^2) \right. \nn \\
&& \times \left. (1+n_3\alpha_{22}+n_4\alpha_{22}^2)i \right],
\label{eq:D_fit1} \\
A(i,a,\alpha_{13},\alpha_{22}) &\approx& (1+A_1\alpha_{13}+A_2\alpha_{13}^2+A_3\alpha_{13}^3) \nn \\
&& \times (1+A_4\alpha_{22}+A_5\alpha_{22}^2+A_6\alpha_{22}^3) \nn \\
&& \times \left[ A_1^{\rm K}(1+A_7\alpha_{13})(1+A_8\alpha_{22}) i^{n_3} \right. \nn \\
&& \left. + A_2^{\rm K}(1+A_9\alpha_{13})(1+A_{10}\alpha_{22}) i^{n_4}\right]^{-1} . \nn \\
\label{eq:A_fit1}
\ea

From these, I obtained the following values for the fit parameters:
\ba
L_1 &\equiv& 0.390575+0.289090a^{3.03136}+0.351485a^{26.0507}, \nn \\
L_2 &\equiv& -0.0210469-0.0704973a^{3.96055}-0.156894a^{30.0636}, \nn \\
L_3 &\equiv& -0.0771662a^{3.30096}-0.112522a^{19.3585}, \nn \\
L_4 &\equiv& -0.0232177a^{2.34745}-0.0474855a^{498.043}, \nn \\
L_5 &\equiv& -0.0910695+0.597861a^{34.2332}, \nn \\
L_6 &\equiv& -0.665419a^{1.47464}, \nn \\
D_1 &\equiv& -0.0471772-0.0714876 a^{4.67811}-2.80254 a^{507.201}, \nn \\
D_2 &\equiv& 0.0100617+0.197316 a^{77.0341}+0.173694a^{194.820}, \nn \\
D_3 &\equiv& 0.175643+0.107683 a^{6.64090}-0.309073 a^{50.4534}, \nn \\
D_4 &\equiv& 0.0144201+0.323182a^{30.8217}-0.377931a^{232.160}, \nn \\
n_1 &\equiv& -0.0617871-0.151930a^{37.0587}+2.21857 a^{131.299}, \nn \\
n_2 &\equiv& 0.00753226-0.00323579 a^{17.8087}+1.40633 a^{97.0069}, \nn \\
n_3 &\equiv& 0.0180080-0.0970060a^{25.9451}+0.439863 a^{104.732}, \nn \\
n_4 &\equiv& -0.0183142-0.228196a^{23.7779}+0.0583285 a^{96.0575}, \nn \\
A_1 &\equiv& 0.0403176-0.458487a^{3.74974}, \nn \\
A_2 &\equiv& -0.0649604+0.0847185a^{1.56286}, \nn \\
A_3 &\equiv& 0.00650092-0.00620249a^{116.004}, \nn \\
A_4 &\equiv& 1.24363+0.211407a^{20.7016}, \nn \\
A_5 &\equiv& 0.281333-0.00508372a^{2.06092}, \nn \\
A_6 &\equiv& 0.0287109-0.266727a^{3.89586}, \nn \\
A_7 &\equiv& 0.286452+0.640543a^{22.2178}, \nn \\
A_8 &\equiv& 0.014822-0.407110a^{6.69441}, \nn \\
A_9 &\equiv& 0.274555-0.722911a^{3.25899}, \nn \\
A_{10} &\equiv& 0.208396+0.799466a^{22.7461}.
\ea

The fit of the diameter is accurate to $<7.5\%$ for spins $a\leq0.95M$ and to $<28.5\%$ for spins $a\leq0.998M$. The fit of the displacement is accurate to $< 13.3\%$ in spin range $0.1M \leq a \leq 0.85M$ with an average accuracy of $2\%$. The accuracy is significantly smaller at high and very low spins and the error can exceed $100\%$ in some cases. Finally, the fit of the asymmetry has an average accuracy of $<4.6\%$, but it can deviate from the simulated data set by factors of order unity, especially at very high spins. If $\alpha_{13}=\alpha_{22}=0$, these fits reduce to the corresponding fits for photon rings around Kerr black holes in Equations~(\ref{eq:L_Kerrfit})--(\ref{eq:A_Kerrfit}).

\section{Conclusions}
\label{sec:conclusions}

In this paper, I analyzed the location of the circular photon orbit and the shapes of photon rings around Kerr-like black holes which are described by a recently proposed metric (Johannsen 2013b). This metric depends on four independent parameters and can serve as a framework to model accretion flows around black holes with arbitrary spin $|a|\leq M$, for which observational signatures of potential violations of the no-hair theorem can then be studied (c.f., Johannsen \& Psaltis 2010a). This work extends the previous analysis of photon rings in JP10b who studied potential violations of the no-hair theorem in terms of a quasi-Kerr metric (Glampedakis \& Babak 2006) up to values of the spin $|a|\lesssim0.4M$. I showed that the radius of the circular photon orbit depends on only two deviation parameters which can shift its location significantly. I derived semi-analytic expressions of the photon ring in the image plane as observed at a large distance away from the black hole and of the inferred diameter, displacement, and asymmetry of the ring. From large sets of simulated ring images, I calculated approximate formulae of the ring diameter, displacement, and asymmetry of both Kerr and Kerr-like black holes. I determined the accuracy of these formulae which is very high for Kerr black holes and moderate for Kerr-like black holes.

In the case of a Kerr black hole, the ring diameter depends only very weakly on the spin and the inclination and can, therefore, be used to improve the mass measurement of Sgr~A* in combination with the monitoring of stars on close orbits around the supermassive black hole (Johannsen et al. 2012). For photon rings in the new Kerr-like metric, I showed that the image size depends primarily on the mass and the deviation parameter $\alpha_{13}$. Therefore, this parameter could be determined from the image size if the mass and distance are known with sufficient accuracy. Currently, Sgr~A* has a measured mass and distance with uncertainties of only $\approx9\%$ and $\approx5\%$, respectively (Gillessen et al. 2009), which are comparable to the changes of the ring diameter introduced by the parameter $\alpha_{13}$. The uncertainty of the mass of the supermassive black hole at the center of M87 is currently reported to be $\approx29\%$ assuming no uncertainty in the distance (see G\"ultekin et al. 2009 and references therein; see, however, Walsh et al. 2013).

As pointed out by JP10b, the displacement and asymmetry of the photon ring only need to be measured in units of the ring diameter in order to infer the black hole spin and deviation from the Kerr metric. An absolute measurement of these quantities, which would be affected by the uncertainties in the mass and distance measurements, is not necessary. While the displacement of the ring depends primarily on the spin of the black hole, a measurement of the spin from the displacement will be very difficult in practice, because the exact position of the black hole in the image would have to be determined. The asymmetry, on the other hand, depends strongly on the deviation parameters. Since photon rings around Kerr black holes are nearly circular except for values of the spin $a\gtrsim0.9M$ and at large inclinations (JP10b; Chan et al. 2013), the ring asymmetry is a direct probe of a violation of the no-hair theorem (c.f., JP10b).

The EHT is expected to achieve an unprecedented angular resolution which lies significantly below the angular sizes of the photon rings of Sgr~A* and the supermassive black hole in M87. Thanks to the planned inclusion of the Atacama Large Millimeter/submillimeter Array (ALMA) and state-of-the-art receivers at the other stations, the EHT will also be highly sensitive. Especially the measurement of closure phases between the stations will allow for a highly-resolved study of the shadows and the photon rings of these sources (c.f., Broderick et al. 2011b). Johannsen et al. (2012) estimated the flux density of the photon ring of Sgr~A* to be on the order of 0.1~Jy which means that the ring should be sufficiently bright to be imaged by the EHT.

In order to assess the prospects of the EHT to detect potential deviations from the Kerr metric, detailed simulations of the photon ring will be required which take into account the actual source flux as well as the resolution and sensitivity of the telescope array. A recent analysis of the current VLBI data in the context of a RIAF accretion flow model in a quasi-Kerr background placed first constraints on such deviations (Broderick et al. 2013). Depending on the magnitude of the deviation parameter, these models also predict different amounts of correlated flux density which should be distinguishable at least along baselines between ALMA and sites in North America (see Figure~5 in Fish et al. 2013).

Here I limited my analysis to include only thin photon rings which delineate the shadow silhouette of a black hole if its accretion flow is optically thin, which is the case at least for Sgr~A* at submillimeter wavelengths. Actual photon rings, however, have a small thickness and the ring images I calculated in this paper coincide with the inner edges of these photon rings, which are marked by a sharp flux fall-off. Further refined models of the ring which take into account its thickness will be required for an image analysis, but the tools developed here should be very useful for such models. Due to the relativistic effects of boosting and beaming, the ring thickness is not uniform and is largest near the equatorial plane on the side of the black hole that approaches the observer. Kamruddin \& Dexter (2013) recently constructed a simple crescent model to incorporate the varying ring thickness (see, also, L. Benkevitch et al., in preparation). 

In practice, the image of the photon ring is slightly blurred by the array resolution and interstellar scattering and needs to be disentangled from the image of the accretion flow in the $uv$-plane, which is only partially sampled by a given set of baselines. Johannsen et al. (2012) used the location of the nulls of the visibility function as a benchmark for the uncertainties in such a measurement and argued that it should be feasible with the EHT. A full analysis will need to extract the ring with a pattern matching technique of the full visibility function.

In the case of Sgr~A*, the no-hair theorem may also be tested with two different techniques in the regime of weak gravitational fields. The presence of the spin of the black hole as well as its quadrupole moment introduces torques on stars in close orbits around Sgr~A* which cause their orbital planes to precess (Will 2008). While this effect is too small for the currently known S-stars in the vicinity of Sgr~A*, because their average distances from the center of mass are too large, these precession frequencies might be measurable for stars within $\sim1000r_S$ of the black hole, if they are monitored over a sufficiently long time (Merritt et al. 2010). For these stars, potentially perturbing effects such as those of other stars (Sadeghian \& Will 2011) and dark matter (Sadeghian et al. 2013), as well as drag forces exerted by the hot plasma surrounding Sgr~A* (Psaltis 2012), may be negligible. Such observations could be obtained using the second generation instrument GRAVITY for the Very Large Telescope Interferometer (Eisenhauer et al. 2011).

In addition, if one of the stars orbiting around Sgr~A* is a pulsar, timing observations may measure characteristic spin-orbit residuals that are induced by the quadrupole moment of the black hole (Wex \& Kopeikin 1999; Pfahl \& Loeb 2004; Liu et al. 2012b). Recently, Mori et al. (2013) reported the discovery of the first pulsar in the vicinity of the Galactic center which lies at a distance of only $~\sim0.1~{\rm pc}$ (Rea et al. 2013) corresponding to $\sim2.5\times10^5r_S$. Future measurements obtained by the monitoring of stars, timing observations of pulsars, and VLBI imaging are largely independent of each other and will be affected by different systematic uncertainties. The combination of their results could lead to a robust test of the no-hair theorem.

\vspace{0.5cm}

This work was supported by a CITA National Fellowship at the University of Waterloo and in part by Perimeter Institute for Theoretical Physics. Research at Perimeter Institute is supported by the Government of Canada through Industry Canada and by the Province of Ontario through the Ministry of Research and Innovation.

\appendix

The set of Equations~(\ref{eq:photonorbit}) has two solutions. One of them corresponds to one single point in the image plane, which is unphysical. Therefore, this solution has to be rejected. The parameters $\xi$ and $\eta$ in Equations~(\ref{eq:xi})--(\ref{eq:eta}) are then given by the expressions
\ba
\xi &=& -\frac{-3 a^4 M^3 \alpha_{13}+a^2 r (7 M^4 \alpha_{13}-5 M^3 r \alpha_{13}+M r^3+r^4)+r^3 (3 M^4 \alpha_{13}-2 M^3 r
   \alpha_{13}-3 M r^3+r^4)}{a r [M^2 \alpha_{22} (2 a^2+3 r^2)-5 M^3 r \alpha_{22}-M r^3+r^4]}, \\
\eta &=& \frac{1}{a^2 r^6 [M^2 \alpha_{22} (2 a^2+3 r^2)-5 M^3 r \alpha_{22}-M r^3+r^4]^2} \big\{
a^8 M^7 \alpha_{13} \alpha_{22} (M^3 \alpha_{13} \alpha_{22}+6 M r^2 \alpha_{13} \nn \\
&& -4 r^3 \alpha_{22})-a^6
   M^4 r \big[2 M^7 \alpha_{13}^2 \alpha_{22}^2+M^6 r \alpha_{13}^2 \alpha_{22}^2+12 M^5 r^2 \alpha_{13}^2 \alpha_{22}-2 M^4 r^3 \alpha_{13} \alpha_{22} (\alpha_{13}+4 \alpha_{22}) \nn \\
&& +8 M^3 r^4 \alpha_{13} (\alpha_{22}^2-3
   \alpha_{13})+M^2 r^5 \alpha_{13} (15 \alpha_{13}+34 \alpha_{22})+6 M r^6 \alpha_{22} (\alpha_{13}-2
   \alpha_{22})-8 r^7 \alpha_{22}^2\big] \nn \\
&& +a^4 M^2 r^3 \big\{4 M^9 \alpha_{13}^2 \alpha_{22}^2-M^8 r \alpha_{13}^2
   \alpha_{22}^2+8 M^7 r^2 \alpha_{13}^2 \alpha_{22}+M^6 r^3 \alpha_{13} \big[8 \alpha_{22}^2-\alpha_{13} (6
   \alpha_{22}+49)\big] \nn \\
&& +2 M^5 r^4 \alpha_{13} \big[38 \alpha_{13}+\alpha_{22} (2 \alpha_{22}+35)\big]-5 M^4 r^5 (6 \alpha_{13}^2+8 \alpha_{13} \alpha_{22}+5 \alpha_{22}^2)+2 M^3 r^6 \alpha_{22} (\alpha_{13}-\alpha_{22}) \nn \\
&& +M^2 r^7
   (23 \alpha_{22}^2-22 \alpha_{13})+2 M r^8 (5 \alpha_{13}+8 \alpha_{22})+8 r^9 \alpha_{22}\big\}+a^2 M r^5
   \big\{-2 M^{10} \alpha_{13}^2 \alpha_{22}^2+M^9 r \alpha_{13}^2 \alpha_{22}^2 \nn \\
&& +4 M^8 r^2 \alpha_{13}^2 \alpha_{22}-2
   M^7 r^3 \alpha_{13} \big[\alpha_{13} (\alpha_{22}+21)+8 \alpha_{22}^2\big]+2 M^6 r^4 \alpha_{13} \big[28\alpha_{13}+\alpha_{22} (4 \alpha_{22}+15)\big] \nn \\
&& -M^5 r^5 \alpha_{13} (19 \alpha_{13}+30 \alpha_{22})+2 M^4 r^6 \big[\alpha_{13}
   (4 \alpha_{22}+21)-16 \alpha_{22}^2\big]-2 M^3 r^7 \big[24 \alpha_{13}+(15-8 \alpha_{22}) \alpha_{22}\big] \nn \\
&& +2 M^2 r^8 (7
   \alpha_{13}-2 \alpha_{22})+10 M r^9 \alpha_{22}+4 r^{10}\big\} - r^{10}(3 M^4 \alpha_{13}-2 M^3 r \alpha_{13}-3
   M r^3+r^4)^2 \big\}.
\ea

In the Kerr limit, these parameters reduce to the familiar expressions (Bardeen 1973)
\ba
\xi &=& -\frac{ r^2(r-3M)+a^2(r+M) }{ a(r-M) }, \\
\eta &=& \frac{ r^3 [4a^2M-r(r-3M)^2] }{ a^2(r-M)^2 }.
\ea

{}

\end{document}